\documentclass[12pt]{article}

\usepackage{indentfirst}
\usepackage[utf8]{inputenc}
\usepackage[T1]{fontenc}
\usepackage{lmodern}
\usepackage{amsmath, amssymb, amsfonts}
\usepackage{xcolor}
\usepackage{graphicx}
\usepackage{caption}
\usepackage{subcaption}
\usepackage{booktabs}
\usepackage{array}
\usepackage{authblk}
\usepackage{titlesec}
\usepackage{setspace}
\usepackage{geometry}
\usepackage{ragged2e}
\usepackage{float}
\usepackage{hyperref}
\usepackage{gensymb}
\usepackage{threeparttable}
\usepackage{amsmath}
\usepackage{arydshln}
\usepackage{graphicx}
\usepackage{booktabs}
\usepackage{multirow}
\makeatletter
\renewcommand{\@seccntformat}[1]{%
  \csname the#1\endcsname\quad}
\makeatother

\graphicspath{{overleaf/}}

\titleformat{\section}{\large\bfseries}{\thesection}{1em}{}

\usepackage{color}  
\definecolor{gray}{gray}{.4}

\newcommand{\URL}[1]{\textcolor{blue}{\url{#1}}}  

\newcommand \MZ [1]{\bgroup\noindent[\textcolor{blue}{\textbf{MZ}: #1}]\egroup\ignorespacesafterend}

\title{\textbf{Multiscale Modelling of Ferroelectrics using a Physics-Informed Neural Network Driven by Molecular Dynamics Data: Parameter Identification and Field Reconstruction}}
\date{}

\author[1,*]{Xuejian Wang}
\author[1]{Frank Wendler}
\author[2]{Hikaru Auzuma}
\author[1]{Michael Zaiser}
\author[2]{Shuji Ogata}
\author[2]{Ryo Kobayashi}
\author[1]{Lei Zeng}
\author[1]{Xingchen Tan}

\affil[1]{Institute of Materials Simulation, Department of Materials Science and Engineering, Friedrich-Alexander-Universität Erlangen-Nürnberg, Dr.-Mack-Str. 77, 90762 Fürth, Germany}
\affil[2]{Graduate School of Engineering, Nagoya Institute of Technology, Nagoya 466-8555, Japan}

\begin{document}

\maketitle

\noindent\textbf{Keywords:} Physics-Informed Neural Network; Multi-scale Model; Phase-Field Method; Molecular Dynamics Simulation; Ferroelectrics

\vspace{1em}
\noindent\textbf{Abstract}

\noindent In multiscale modeling of ferroelectric materials, combining atomistic simulation with continuum-scale phase-field model (PFM) remains a fundamental challenge. One of the key difficulties lies in faithfully capturing discrete atomic-level information within a continuum modeling framework, while accurately representing material behavior at the mesoscale. In this paper, a Physics-Informed Neural Network (PINN) framework driven by molecular dynamics (MD) data is developed. The loss function of the network consists of two components: a supervised term that fits the discrete spatial polarization distributions obtained from MD simulations of systems containing domain walls, and a physics-based term that incorporates the residuals of partial differential equations (PDEs) of steady-state PFM. To ensure stable and balanced training among the different physical loss components, an adaptive gradient normalization (GradNorm) strategy is employed to dynamically adjust the task weights based on their gradient magnitudes. By minimizing the total loss, the model not only reconstructs the polarization field along with the associated strain, stress, and energy landscape at the continuum scale, but also identifies critical physical parameters required for the phase-field model, including the characteristic energy density, characteristic length factor, gradient energy anisotropy factor, and Landau polynomial coefficients. By using the PINN-predicted physical parameters in COMSOL Multiphysics to solve the corresponding PDEs within a finite element framework, we demonstrate that the learned parameters enable accurate reproduction of not only the ferroelectric domain structure, but also the associated material response, including stress/strain distributions and the energy landscape. This framework provides an effective methodology for establishing multiscale connections between atomistic and continuum descriptions, and holds the potential to infer underlying physical properties directly from polarization distributions for a wide range of materials.

\section{Introduction}
Ferroelectric materials have garnered significant attention in condensed matter physics and materials science due to their remarkable physical properties, including piezoelectricity \cite{zhang2024te}, pyroelectricity \cite{zhao2025why}, electro-optic effects \cite{thapa2024microsecond}, and nonlinear dielectric response \cite{zhao2024dielectric}. These unique characteristics are responsible for a wide array of applications in advanced technologies such as sensors, actuators, non-volatile memories, and energy conversion and storage devices. Among ferroelectric materials, Barium Titanate (BaTiO$_3$, BTO) represents a prototypical model system with perovskite structure and well-known properties \cite{liu2025flexible, liu2025superior, yang2025effects}. The excellent ferroelectric and piezoelectric properties of this lead-free perovskite material make BTO an environmentally benign and cost-effective alternative to traditional lead-based ferroelectrics such as lead zirconate titanate (PZT). 

The remarkable functionalities of ferroelectric materials originate from their intrinsic spontaneous polarization, which self-organizes into domains separated by domain walls (DWs). These DWs -- interfaces between regions of differing polarization orientation -- are, in essence, structural and energetic defects within the crystal \cite{Catalan2012DomainWallNanoelectronics, Nataf2020DomainWallEngineering,Chen2023TopologicalPhononic, zheng2025chiral}. As such, they introduce local discontinuities in polarization and significant concomitant perturbations in electrical and mechanical fields. These localized disturbances, while confined at the nanoscale, play an essential role because they control the macroscopic electromechanical and even photo-electrical properties of the material. For instance, Matsuo~\cite{matsuo2023} demonstrated that the anomalous photovoltaic response in ferroelectric materials originates from the built-in electric field ($E_\mathrm{local}$) at domain walls. Concerning mechanical behavior, one of the well-known examples is the domain wall pinning effect. A large body of research has focused on enhancing the stability of domain walls, often referred to as ``hardening'' the domain structure. Strategies to achieve this include, but are not limited to, the introduction of defects (e.g., through doping or dislocations)~\cite{gao2023topology,hofling2021control}, coherent plate-like precipitates~\cite{zhao2021precipitation,zhao2022coherent}, and optimization of fabrication methods~\cite{tao2024}. In this regard, Tao et al.~\cite{tao2024} reported the preparation of lead-free (K,Na)NbO$_3$ (KNN) piezoceramics via hot-pressing (HP). Compared to conventionally sintered samples, the HP-KNN ceramics exhibit significantly improved dielectric permittivity, reduced dielectric loss, and enhanced thermal stability. These enhancements are closely linked to the local stress and strain fields near domain walls. Therefore, gaining deep insight into the local strain/stress fields and built-in electric fields at domain walls is of critical importance for modulating the electromechanical response and overall functional performance of ferroelectric materials.

To experimentally probe these localized fields in bulk ferroelectric materials, a variety of advanced characterization techniques have been developed. For built-in electric fields, high-resolution techniques such as Kelvin Probe Force Microscopy (KPFM)~\cite{maguire2024direct,doherty2023domain} and off-axis Electron Holography have been employed to map electrostatic potential variations at domain walls. These methods provide insights into the local electric field distribution and potential steps associated with domain wall configurations. Regarding local stress and strain fields, synchrotron-based nanofocused X-ray diffraction (nano-XRD) techniques are powerful tools to non-destructively quantify three-dimensional strain distributions in the vicinity of ferroelectric domain walls~\cite{hadjimichael2018domain}. These complementary methods offer valuable insights into the coupled electromechanical microenvironments that govern macroscopic ferroelectric performance. While the mentioned methods are indeed capable of probing domain wall fields in bulk ferroelectric materials, their practical applicability is often limited by several factors. These include the need for expensive, large-scale instrumentation, stringent sample preparation protocols, and the restriction of spatial resolution, especially in a surface-normal direction. Moreover, such methods typically provide only static snapshots of material states and are less suited for capturing dynamic domain evolution under external stimuli. These challenges highlight the growing importance of computational modeling as a complementary tool for investigating domain wall phenomena with high spatial and temporal resolution.

Given the experimental hurdles, computational modeling emerges as a vital complementary tool for exploring domain wall phenomena with higher spatial and temporal resolution. On the computational side, methods ranging from first-principles calculations based on density functional theory (DFT) to continuum-scale simulations offer complementary perspectives \cite{liu2021quadrupole,zhou2025cooling,mi2021breakdown,liu2022vortex,Zhu2024HfO2Review,Ali2025PbTiO3Adsorption,Zhou2026ElectromechanicalNBT}. Among these, multi-scale modeling approaches that bridge information across different spatial and temporal scales are particularly valuable for achieving a comprehensive understanding of domain wall behavior. As a classical mechanics-based method, molecular dynamics (MD) provides atomistic insight into the dynamic behavior of materials by explicitly solving Newton's equations of motion for every atom in the system. It captures detailed atomic information such as positions, velocities, local temperatures, and electrostatic potentials. Through statistical analysis of atomic configurations, MD allows accurate characterization of domain wall width and energy per unit area across different geometrical and topological configurations. Notably, MD simulations can realistically capture the structural evolution of ferroelectric materials, as they allow both shape and volume relaxation during domain wall formation. Such mechanical freedom enables the system to reach physically consistent polarization and strain states, reflecting the balance between internal and external stresses. In contrast, by solving a set of coupled partial differential equations-including the time-dependent Ginzburg-Landau (TDGL) equation for polarization evolution, mechanical and electrostatic equilibrium equations, the phase-field method (PFM) predicts not only polarization distributions but also the accompanying mechanical and electrostatic fields. The use of Fourier transform or finite element methods to minimize the total free energy makes PFM particularly effective for resolving mesoscale features such as the distributions of strain/stress, built-in electric fields, and energy densities.

Consequently, integrating molecular dynamics (MD) and phase‐field modeling (PFM) within a unified multi-scale framework offers significant advantages for characterizing ferroelectric domain wall behavior, provided that the underlying interatomic potentials are accurately calibrated. A key bottleneck, however, lies in reliably extracting the parameters required by PFM directly from atomistic simulations. These include the characteristic energy density scale ($G$), characteristic domain‐wall length scale ($l$), gradient‐energy anisotropy ($\mu$), and Landau polynomial coefficients. Obtaining these quantities from MD is nontrivial: many of them are not directly accessible and typically require fitting procedures, simplifying assumptions, or intermediate continuum models, complicating seamless MD–PFM coupling.

To partially address this challenge, our previous work~\cite{durdiev2025parameterization} introduced a systematic MD-driven parameterization strategy for BaTiO$_3$. That study extracted elastic and piezoelectric tensors, Landau coefficients, DW energies and widths, and kinetic parameters from MD, and incorporated them into a fully anisotropic PF formulation. The resulting PFM quantitatively reproduced MD predictions of 180° and 90° DW structures, velocities, and coercive fields across different thermodynamic ensembles, establishing an atomistic-to-mesoscale bridge. Despite this important progress, several key limitations remain. The parameter extraction procedure is still computationally intensive, relies on computationally demanding and specifically tailored MD simulations, and requires manual calibration steps to ensure consistency across energy landscapes and boundary conditions.

Physics-informed neural networks (PINN)~\cite{karniadakis2021physics, cuomo2022scientific, raissi2020hidden, rojas2023parameter, Chen2025PINNJoint, Shang2024GradientPINN} have demonstrated to be a viable approach across various fields and are poised to become powerful tools for addressing the aforementioned difficulties. Specifically, PINNs offer a unique advantage by seamlessly integrating governing physical laws into their architecture, thereby inherently enforcing physical consistency in their predictions. Moreover, PINNs provide direct access to analytical derivatives of the network outputs through automatic differentiation. This is particularly beneficial for phase-field models, where the PDE residuals involve high-order spatial derivatives of polarization (e.g., gradient-energy terms), and it avoids additional discretization errors that may arise from finite-difference stencils or the choice of finite-element interpolation. For instance, Raissi et al.~\cite{raissi2019physics} proposed a general PINN framework capable of solving both forward and inverse problems governed by nonlinear partial differential equations. In the context of ferroelectric phase-field modeling, Shang et al.~\cite{shang2024quantification} successfully developed a PINN-based inverse modeling approach that uses experimental data to identify gradient energy coefficients. This work represents a significant step toward quantitatively linking mesoscopic models with experimental observables and demonstrates the potential of PINNs in parameter identification within ferroelectric systems. 

Building upon this capability, the present work develops an extended PINN framework that integrates MD data as physical constraints within the loss function. The core idea is to construct a deep neural network, implemented in TensorFlow (TensorFlow v1, with a fully connected residual network backbone), in which the governing partial differential equations of the ferroelectric phase-field model are implicitly embedded as physical constraints within the loss function. Automatic differentiation in TensorFlow is used to evaluate the required spatial derivatives of the network outputs, enabling a straightforward construction of the phase-field PDE residuals. Meanwhile, discrete polarization fields corresponding to steady-state domain wall structures, including 180° and 90° DWs as obtained from atomistic MD simulations, are incorporated as supervised data during the training process. The overall training objective consists of multiple heterogeneous loss components, including data-fidelity terms, Ginzburg–Landau energy residuals, electrostatic and mechanical balance constraints, and geometric regularization terms. This naturally leads to a multi-task optimization problem, where different loss terms correspond to distinct physical tasks with potentially different scales and convergence rates. To address this issue, the GradNorm (GN) method~\cite{chen2018gradnorm} is incorporated as an adaptive loss-balancing strategy. GN dynamically adjusts the relative weights of different loss terms by equalizing their gradient magnitudes with respect to the shared network parameters, thereby promoting balanced training across all physical constraints and preventing domination by a single loss component. Through this end-to-end multi-task learning process, the proposed approach not only solves the coupled partial differential equations of the ferroelectric PFM, but also enables the simultaneous and automatic identification of multiple critical material parameters. In contrast to previous work that focused solely on estimating gradient energy coefficients~\cite{shang2024quantification}, our model extends this capability to include the identification of Landau coefficients, characteristic energy density ($G$), and domain wall width factor ($l$), thereby offering a more comprehensive and versatile inverse modeling framework for ferroelectric systems. Furthermore, it can also predict all other physical fields associated with the steady-state domain wall structures, including strain fields, electric fields, electric displacement fields, and the distributions of different energy terms. This provides a unified route to determine complex parameters and integrate data from MD and PFM simulations.

The remainder of this paper is organized as follows. Section ~\ref{sec:tf} presents the overall workflow of this paper, from MD data preprocessing to the structure of the considered PFM and the architecture of the PINN, along with PFM implementation details within a FEM framework. Section~\ref{sec:rd} provides a detailed presentation of the simulation results and offers an in-depth discussion of their physical significance.

\section{Theoretical Framework}
\label{sec:tf}

\subsection{Workflow}
\label{sec:workflow}

\begin{figure}[H]
    \centering
    \includegraphics[width=0.35\textwidth]{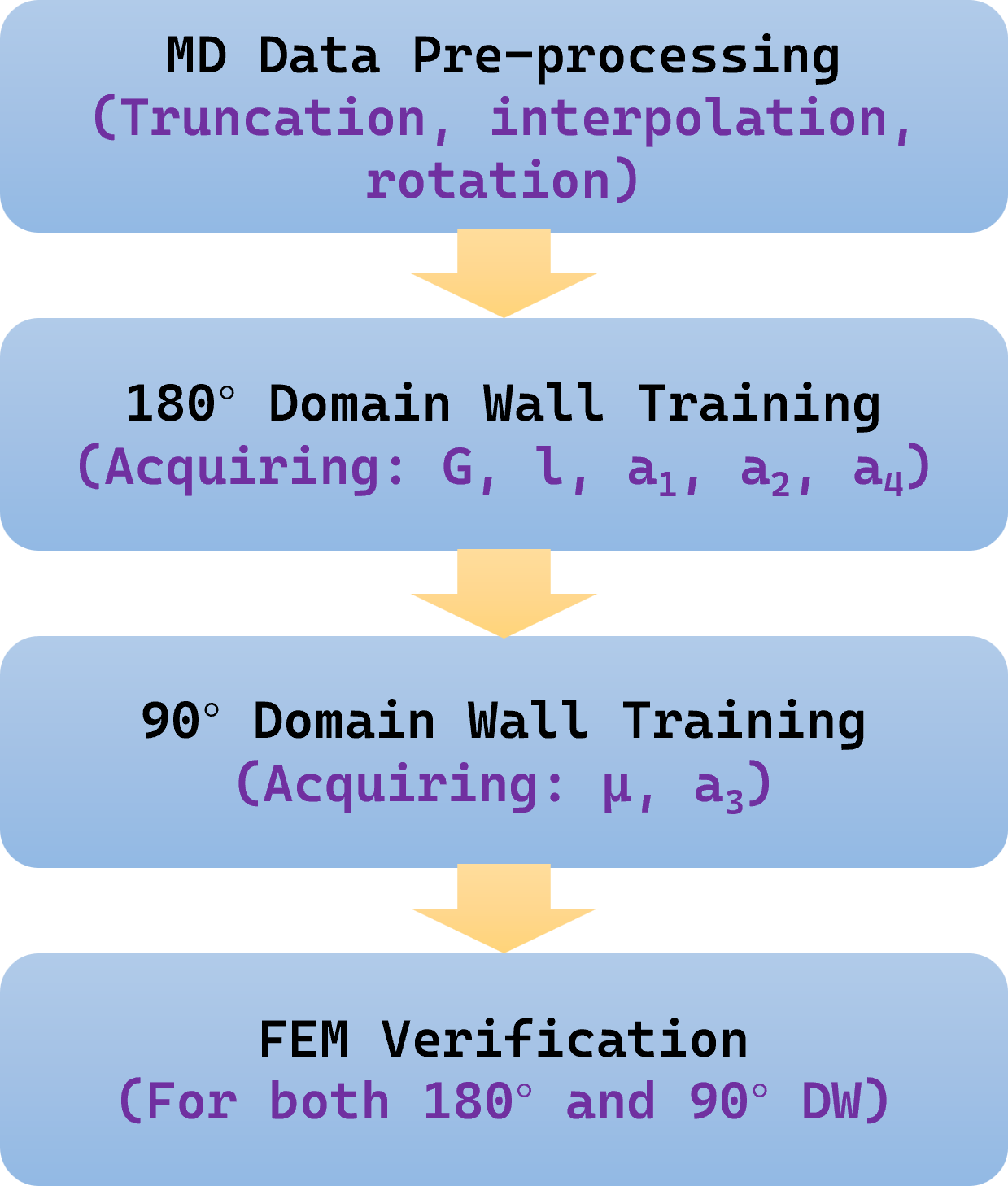}
    \caption{Workflow for PINN training of MD-data driven ferroelectric phase-field model}
    \label{fig:workflow}
\end{figure}

This section outlines the principal workflow of this study, as illustrated in Fig.~\ref{fig:workflow}. The process begins with the acquisition of the discrete polarization distribution from MD simulations, that include the cases of stable 180° and 90° DW configurations. Before being used for neural network training, the MD data is subjected to several preprocessing steps—including spatial truncation, coordinate normalization, and rotational alignment—to ensure numerical consistency and physical correctness. Data truncation removes the data from atoms near the periodic boundaries, where atomic positions may wrap across opposite faces of the simulation cell under periodic boundary conditions, leading to artificial discontinuities in the extracted fields. Coordinate normalization rescales spatial coordinates to a dimensionless form, which stabilizes neural‐network training and accelerates convergence. Coordinate system rotation is applied specifically to the 90° domain wall dataset. In this case, the MD simulation domain is a slab where the periodic boundaries necessitate a rotation by 45° relative to the Cartesian axes (with $x$, $y$, and $z$ aligned to the [100], [010], and [001] crystallographic directions)~\cite{azuma2023microscopic}. This requires a consistent transformation to map the MD frame to the phase‐field coordinate system. These steps are described in detail in ~\ref{app:md_preprocessing}. 

The processed data are subsequently employed for two sequential training stages within the PINN framework described in Section~\ref{sec:pinn}. In the first stage, the MD-derived 180$^\circ$ DW profile serves as supervised input, enabling the learning of the Landau coefficients $a_1$, $a_2$, and $a_4$, together with the characteristic energy density factor $G$ and the characteristic DW width factor $l$. In the second stage, the MD 90$^\circ$ DW profile is used for supervising the training process while the parameters learned in the first stage are fixed; this step yields the anisotropy factor $\mu$ and the remaining Landau coefficient $a_3$, which have an effect only in the presence of at least two non-zero polarization components and are therefore relevant exclusively for the 90$^\circ$ DW. Each stage of training reconstructs not only the target polarization field but also the associated strain, stress, as well as the different PFM energy density contributions for the corresponding DW configurations.

Finally, the full parameter set obtained from the multi-stage PINN training is used in the FEM framework to perform two numerical simulations following the procedure outlined in Section~\ref{sec:fem}. The FEM-computed polarization, strain, stress, electric field, and energy density distributions are then compared with the PINN-reconstructed fields and with the MD-derived polarization data, providing a quantitative validation of the PINN framework.

\subsection{Phase-field Model}
\label{sec:pfm}

\subsubsection{Constitutive equations}

\begin{figure}[H]
    \centering
    \includegraphics[width=0.45\textwidth]{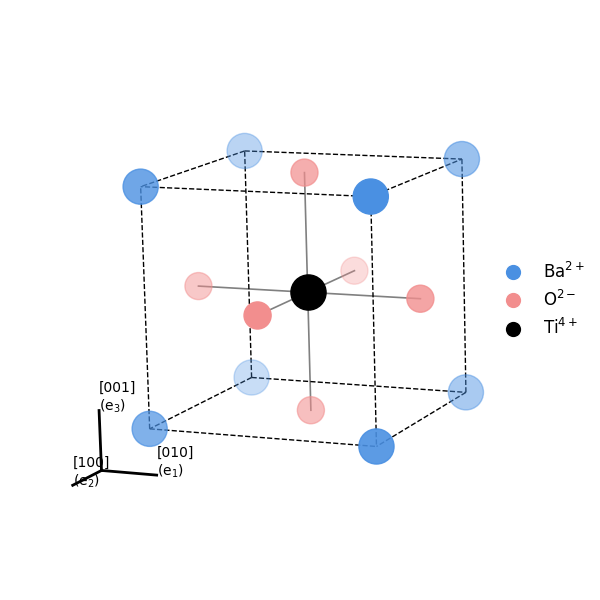}
    \caption{Crystal Structure of BTO}
    \label{fig:LatticeStructure}
\end{figure}

We adopt the continuum PFM proposed by Durdiev et al.~\cite{durdiev2025parameterization}, in which the polarization field $\mathbf{P}$ is taken as the primary order parameter. Its evolution is governed by the time-dependent Ginzburg--Landau (TDGL) equation
\begin{equation}
\beta \frac{\partial \mathbf{P}}{\partial t}
=
- \frac{\delta \mathcal{H}}{\delta \mathbf{P}},
\qquad
\mathcal{H}
=
\int H(\mathbf{P}, \mathbf{E}, \boldsymbol{\varepsilon}) \, d^3\mathbf{r},
\label{eq:TDGL_1}
\end{equation}
where $\mathcal{H}$ is the total (free) enthalpy functional, $H$ is the electrical enthalpy density, $\mathbf{E}$ denotes the electric field, and $\boldsymbol{\varepsilon}$ is the total strain. $\beta$ is a positive-definite mobility coefficient controlling the domain wall (DW) mobility. 

In the ferroelectric phase, the spontaneous polarization can orient along three possible, symmetry equivalent axes, marked by the unit vectors $\{\mathbf{e}_i\}$. These vectors define the crystallographic reference directions of the cubic unit cell illustrated in Fig.~\ref{fig:LatticeStructure}. Accordingly, the energy functional is formulated with reference to a cubic lattice structure in which the three possible spontaneous polarization directions are equivalent. The tetragonal symmetry of the ferroelectric phase then emerges from symmetry breaking as the energy functional is constructed to possess, in absence of external fields, six equivalent minima defining the possible spontaneous polarization values. 

Following Ref.~\cite{durdiev2025parameterization}, also the mobility coefficient $\beta$ is taken to possess cubic symmetry
\begin{equation}
\beta^{-1}
= \beta_0^{-1}
\left[
1 - \eta \left(1 - \sum_i (\mathbf{e}_i \cdot \mathbf{e}_H)^4 \right)
\right],
\end{equation}
where $\eta$ is the anisotropy coefficient. Here, $\mathbf{e}_H$ denotes the unit vector along the thermodynamic driving force:
\begin{equation}
\mathbf{e}_H
=
\frac{\delta \mathcal{H}/\delta \mathbf{P}}
{\left\lVert \delta \mathcal{H}/\delta \mathbf{P} \right\rVert},
\end{equation}
which specifies the local direction of polarization evolution. 

The enthalpy density $H$ in Eq.~(\ref{eq:TDGL_1}) is composed of the domain-separation, gradient, and bulk energy densities, $H = H^{\text{sep}}(\mathbf{P}) + H^{\text{grad}}(\nabla \mathbf{P}) +  H^{\mathrm{bulk}}(\boldsymbol{\varepsilon}, \mathbf{E}, \mathbf{P})$. These are given by

\begin{equation}
    H^{\text{sep}}(\mathbf{P}, \{\mathbf{e}_i\})
    = G\,\psi(\mathbf{P}, \{\mathbf{e}_i\}),
\label{eq:Landau}
\end{equation}

\begin{equation}
    H^{\text{grad}}(\nabla \mathbf{P})
    = \frac{1}{2}\frac{G}{P_s^2}
      (\nabla \mathbf{P}) : \mathbb{L} : (\nabla \mathbf{P}),
\label{eq:GradEnergy}
\end{equation}

\begin{equation}
\begin{aligned}
H^{\mathrm{bulk}}(\boldsymbol{\varepsilon}, \mathbf{E}, \mathbf{P})
=&\;
\frac{1}{2}\,(\boldsymbol{\varepsilon}^{t}-\boldsymbol{\varepsilon}^{sp})
:\mathbb{C}:
(\boldsymbol{\varepsilon}^{t}-\boldsymbol{\varepsilon}^{sp})
\\
&\;-
(\boldsymbol{\varepsilon}^{t}-\boldsymbol{\varepsilon}^{sp})
:\bigl(\mathbb{E}\cdot \mathbf{E}\bigr)
\\
&\;-
\frac{1}{2}\,\mathbf{E}\cdot\bigl(\mathbb{K}\cdot \mathbf{E}\bigr)
-
\mathbf{P}\cdot \mathbf{E}.
\end{aligned}
\label{eq:BulkEnergyDensity}
\end{equation}

The domain separation energy in Eq.~\eqref{eq:Landau} is defined as the product of a characteristic energy parameter $G$ and a normalized sixth-order Landau polynomial
\begin{equation}
\psi(\mathbf{P}, \{\mathbf{e}_i\}) =
\frac{a_1}{P_s^2} \sum_i (\mathbf{P} \cdot \mathbf{e}_i)^2
+ \frac{a_2}{P_s^4} \sum_i (\mathbf{P} \cdot \mathbf{e}_i)^4
+ \frac{a_3}{P_s^4} \sum_{i \neq j} (\mathbf{P} \cdot \mathbf{e}_i)^2 (\mathbf{P} \cdot \mathbf{e}_j)^2
+ \frac{a_4}{P_s^6} \sum_i (\mathbf{P} \cdot \mathbf{e}_i)^6.
\label{eq:landauPoly}
\end{equation}
where $a_1$, $a_2$, $a_3$, and $a_4$ are the Landau coefficients. Here, ``normalized'' means that the polynomial is formulated in dimensionless form by scaling the polarization components of $\mathbf{P}$ with respect to the spontaneous polarization magnitude $P_s$, while the absolute energy scale is introduced separately through the prefactor $G$. Following Ref.~\cite{durdiev2025parameterization}, the polynomial is constrained to reproduce a double-well energy landscape with minima located at $\mathbf{P}_j = \pm P_s \mathbf{e}_j$. This is achieved by imposing the conditions

\begin{equation}
\psi(\mathbf{0}) = 0, \qquad
\psi(P_s \mathbf{e}_j) = -1, \qquad
\left.\frac{\partial \psi}{\partial P_j}\right|_{\mathbf{P}=P_s\mathbf{e}_j} = 0,
\end{equation}

which yield the relations

\begin{equation}
a_1 + a_2 + a_4 = -1, \qquad
a_1 + 2a_2 + 3a_4 = 0.
\label{eq:a124relationship}
\end{equation}

Accordingly, $a_1$ and $a_2$ can be expressed in terms of $a_4$ as

\begin{equation}
a_1 = a_4 - 2, \qquad
a_2 = 1 - 2a_4.
\label{eq:a124}
\end{equation}

As a result, only $a_4$ remains as an independent coefficient controlling the one-dimensional double-well profile, whereas $a_3>0$ governs the energy penalty associated with mixed polarization components across different polarization variants in both $180^\circ$ and $90^\circ$ DWs. Consequently, $\psi$ characterizes the topology of the energy landscape, i.e., the relative heights and locations of the energy barriers between different variants, while the absolute energy scale is determined by the prefactor $G$.

The gradient energy density (Eq.~\ref{eq:GradEnergy}) penalizes spatial variations in the polarization field and thus promotes smooth polarization transitions across the domain wall. We write this energy density in terms of the characteristic energy parameter $G$ used also as prefactor of the Landau polynomial, and a rank-four length scale tensor ($\mathbb{L}$) which is given by

\begin{equation}
\mathbb{L} = l^2 \left[ \mathbb{I} + \mathbf{I} \otimes \mathbf{I} + \mu \sum_i \mathbf{e}_i \otimes \mathbf{e}_i \otimes \mathbf{e}_i \otimes \mathbf{e}_i \right],
\label{eq:anisotropic_tensor}
\end{equation}

where $l$ is a characteristic domain wall width parameter, and $\mu$ characterizes the strength of the cubic anisotropy governing the 90$^\circ$ and 180$^\circ$ DW energies. 

As shown in Eq.~(\ref{eq:BulkEnergyDensity}), the bulk enthalpy 
of the system consists of 4 contributions: elastic energy density, piezoelectric coupling, electrostatic energy density and dipole energy. The elastic energy density represents the mechanical energy stored due to lattice deformation. It accounts for the mismatch between the total strain $\boldsymbol{\varepsilon}^{t}$ and the polarization-induced eigenstrain. Relative to the cubic reference configuration introduced above, the latter is assumed to be volume-preserving (trace-free) and to take the deviatoric form~\cite{schrade2014invariant}

\begin{equation}
\boldsymbol{\varepsilon}^{sp}(\mathbf{P}) = \frac{3}{2} \frac{\varepsilon_{0}}{P_{s}^{2}}
\left( \mathbf{P} \otimes \mathbf{P} - \frac{1}{3} \lVert \mathbf{P} \rVert^{2} \mathbf{I} \right),
\label{eq:spstrain}
\end{equation}

where $\varepsilon_0$ is the characteristic eigenstrain amplitude associated with the tetragonal ferroelectric distortion. It is defined from the tetragonal lattice parameters $a$ and $c$ of the ferroelectric phase, such that the resulting spontaneous strain preserves, for $\mathbf{P} = \mathbf{P}_j$, the volume $V = ca^2$ of the unit cell

\begin{equation}
    \varepsilon_0
    = 2\,\frac{c-a}{c+2a} \approx 0.781\%.
\end{equation}

The electromechanical coupling energy density captures the interaction between strain and electric field, characteristic of piezoelectric behavior. The influence of electric field induced strain involves the piezoelectric tensor $\mathbb{E}$, which we express as

\begin{equation}
\mathbb{E}(\mathbf{P}) 
= \frac{e_{33} - e_{31} - e_{15}}{P_s^3} \, \mathbf{P} \otimes \mathbf{P} \otimes \mathbf{P}
+ \frac{\|\mathbf{P}\|^2}{P_s^3} \left[ 
e_{31} \, \mathbf{P} \otimes \mathbf{I} 
+ \frac{e_{15}}{2} \left( \mathbf{I} \otimes \mathbf{P} + (\mathbf{P} \otimes \mathbf{I})^{T} \right)
\right],
\end{equation}

where $e_{33}$, $e_{31}$, and $e_{15}$ are the independent piezoelectric coefficients of tetragonal BaTiO$_3$ in the crystal frame, corresponding to the longitudinal, transverse, and shear electromechanical couplings with respect to the polar axis, respectively \cite{Shu2001DomainPatternsMacroscopic}.

The electrostatic and dipole energy density in Eq.~(\ref{eq:BulkEnergyDensity}) consists of two contributions: the stored electric energy in the dielectric medium, $\frac{1}{2}\mathbf{E}\cdot(\mathbb{K}\cdot\mathbf{E})$, and the interaction energy between the electric field $\mathbf{E}$ and the polarization vector $\mathbf{P}$. In contrast to previous work, where an isotropic dielectric tensor $\mathbb{K}$ was assumed~\cite{durdiev2025parameterization,Durdiev2023FourierPF,Kumar2023FerroX,Liu2022VortexChirality,Zhou2022DislocationDW}, the present study adopts an anisotropic dielectric description informed by recent MD results reported by Azuma et al.~\cite{azuma2025unique}, which indicate different dielectric responses parallel and perpendicular to the polarization direction. Accordingly, the dielectric tensor is expressed as

\begin{equation}
\mathbf{K}
= k_a \mathbf{I}
+ (k_c - k_a)\,\mathbf{e}_P\otimes\mathbf{e}_P,
\qquad
\mathbf{e}_P=\frac{\mathbf{P}}{|\mathbf{P}|}.
\label{eq:anisotropic_K}
\end{equation}

where $k_a$ and $k_c$ denote the dielectric coefficients perpendicular and parallel to the local polarization direction, respectively, and $\mathbf{I}$ is the identity tensor.

In the present phase-field formulation, the polarization field $\mathbf{P}$ is explicitly resolved as the ferroelectric soft-mode order parameter. Consequently, the dielectric response associated with this soft mode is already accounted for through the evolution of $\mathbf{P}$ and should not be included again in the dielectric tensor entering the electrostatic energy. However, the dielectric susceptibilities extracted from the MD simulations correspond to the total dielectric response, comprising both the fast background contribution and the soft-mode contribution. To avoid double counting of the soft-mode response, the total susceptibilities are therefore separated into background and soft-mode parts using a temperature-dependent fitting procedure, as described in detail in \ref{app:md_constants}.

From the energy functional given above we can find electrical and mechanical constitutive equations that govern the coupling between polarization, electric field, and mechanical deformation, as

\begin{equation}
\begin{aligned}
\boldsymbol{\sigma} &= \mathbf{C} : (\boldsymbol{\varepsilon} - \boldsymbol{\varepsilon}^{sp}) - \mathbb{E}^\mathrm{T} \cdot \mathbf{E}, \\
\mathbf{D} &= \mathbb{E} : (\boldsymbol{\varepsilon} - \boldsymbol{\varepsilon}^{sp}) + \mathbb{K} \cdot \mathbf{E} + \mathbf{P}.
\end{aligned}
\label{eq:sepEnergy}
\end{equation}

Here, $\boldsymbol{\varepsilon}$ denotes the total strain tensor and $\mathbf{E}$ is the electric field, which are obtained from mechanical displacement $\mathbf{u}$ and the electric potential $\phi$, through
\begin{equation}
\boldsymbol{\varepsilon} = \frac{1}{2}\left( \nabla \mathbf{u} + (\nabla \mathbf{u})^{\mathrm{T}} \right),
\qquad
\mathbf{E} = - \nabla \phi.
\label{eq:strain_electricfield}
\end{equation}

In the present phase-field formulation, the polarization field $\mathbf{P}$ is explicitly resolved as the ferroelectric soft-mode order parameter. Consequently, the dielectric response associated with this soft mode is already accounted for through the evolution of $\mathbf{P}$ and should not be included again in the dielectric tensor entering the electrostatic energy. However, the dielectric susceptibilities extracted from the MD simulations correspond to the total dielectric response, comprising both the fast background contribution and the soft-mode contribution. To avoid double counting of the soft-mode response, the total susceptibilities are therefore separated into background and soft-mode parts using a temperature-dependent fitting procedure, as described in detail in \ref{app:md_constants}.

Assuming that the DWs form or relax during the free evolution of a bulk material without external loads, fields or free charges, the mechanical and electrostatic equilibrium conditions are
\begin{equation}
\mathrm{div}\, \boldsymbol{\sigma} = 0 
\qquad \mathrm{div}\, \mathbf{D} = 0
\end{equation}

Finally, the evolution of the polarization given by Eq.~(\ref{eq:TDGL_1}) can be written as
\begin{equation}
\beta \frac{\partial \mathbf{P}}{\partial t}
= \frac{G}{P_s^2} \nabla \cdot \mathbb{L} : (\nabla \mathbf{P})
- G \frac{\partial \psi(\mathbf{P})}{\partial \mathbf{P}}
- \frac{\partial H^{\mathrm{bulk}}}{\partial \mathbf{P}}.
\label{eq:TDGL_expanded}
\end{equation}

Among the parameters involved in the above equations, the coefficients $G$, $l$, and $a_1$--$a_4$ are the target quantities to be inferred in this work. These coefficients are specific to the phase-field formulation and are not directly available as MD inputs; therefore, the PINN predicts them under MD-informed supervision/constraints. The remaining parameters, including the spontaneous polarization magnitude $P_s$, the elastic stiffness tensor $\mathbb{C}$, the dielectric tensor $\mathbb{K}$, and the lattice constants $a$ and $c$, are directly obtained from MD simulations. These parameters, that can be determined with high accuracy, are listed in Table~\ref{tab:parameters} \cite{azuma2025unique}. Appendix B provides detailed procedures for calculating these parameters in MD.

It is worth noting that the PFM formulation used here, as in most other PF based studies, assumes cubic elastic symmetry. Therefore, the elastic stiffness tensor calculated from MD for the tetragonal phase must be consistently projected onto its effective cubic representation before being incorporated into the PFM framework. The effective cubic elastic constants are evaluated using the relations~\cite{gazis1963elastic,moakher2006closest,ranganathan2008universal} 
\begin{equation}
\begin{aligned}
C_{11} &= \frac{2C_{11}^{T} + C_{33}^{T}}{3} = 173~\text{GPa} \\
C_{12} &= \frac{C_{12}^{T} + 2C_{13}^{T}}{3} = 103~\text{GPa} \\
C_{44} &= \frac{2C_{44}^{T} + C_{66}^{T}}{2} = 80~\text{GPa}
\end{aligned}
\label{eq:cubic_projection}
\end{equation}
with the tensor coefficients from Table \ref{tab:parameters}. 
After applying the cubic projection, the resulting shear modulus $C_{44}$ becomes, however, substantially larger than typical values for cubic BaTiO$_3$. Extensive theoretical calculations and atomistic simulations, including first-principles studies \cite{Wang2010BTO_JAP,Pandech2015ATiO3,Sakhya2015ATiO3} and MD simulations \cite{Hashimoto2015MDpermittivity,Choithrani2014BTOElastic}, consistently report cubic-phase shear moduli on the order of a few tens of gigapascals only. The comparatively large value obtained after the cubic  projection is therefore more reasonably interpreted as a remnant of the tetragonal shear response, rather than a physically admissible cubic stiffness within a cubic phase-field model. To maintain internal consistency with the assumption of cubic symmetry and to stay within the range provided by these studies, we follow the same practical approach  as in our previous work ~\cite{durdiev2025parameterization} and assign $C_{44}$ using the commonly used cubic relation
\begin{equation}
C_{44} = \frac{C_{11} - C_{12}}{2} = 45~\mathrm{GPa}
\end{equation}

\subsubsection{Domain walls}
\label{sec:DWs}

Spatial regions where the polarization $\mathbf{P}$ is close to one of its equilibrium values $\mathbf{P} = \pm P_s \mathbf{e}_j$ are called domains. Boundaries between such regions, where the polarization transitions from one equilibrium state to another, are referred to as domain walls. Domain walls play a key role in the microstructure and functional response of ferroelectric materials, as an important part of the response to electrical and mechanical fields arises from their motion. Accordingly, domain walls can be considered the elementary 'building blocks' of the ferroelectric domain microstructure and the correct representation of their structure and dynamics is the essential criterion for the physical correctness of a phase-field model. For this reason, we focus on domain-wall type solutions in our PINN training programme. 

\begin{figure}[H]
    \centering

    \begin{subfigure}[b]{0.4\textwidth}
        \centering
        \includegraphics[width=\linewidth]{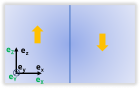}
        \caption{}
        \label{fig:180dw}
    \end{subfigure}
    \hspace{0.01\textwidth}
    \begin{subfigure}[b]{0.4\textwidth}
        \centering
        \includegraphics[width=\linewidth]{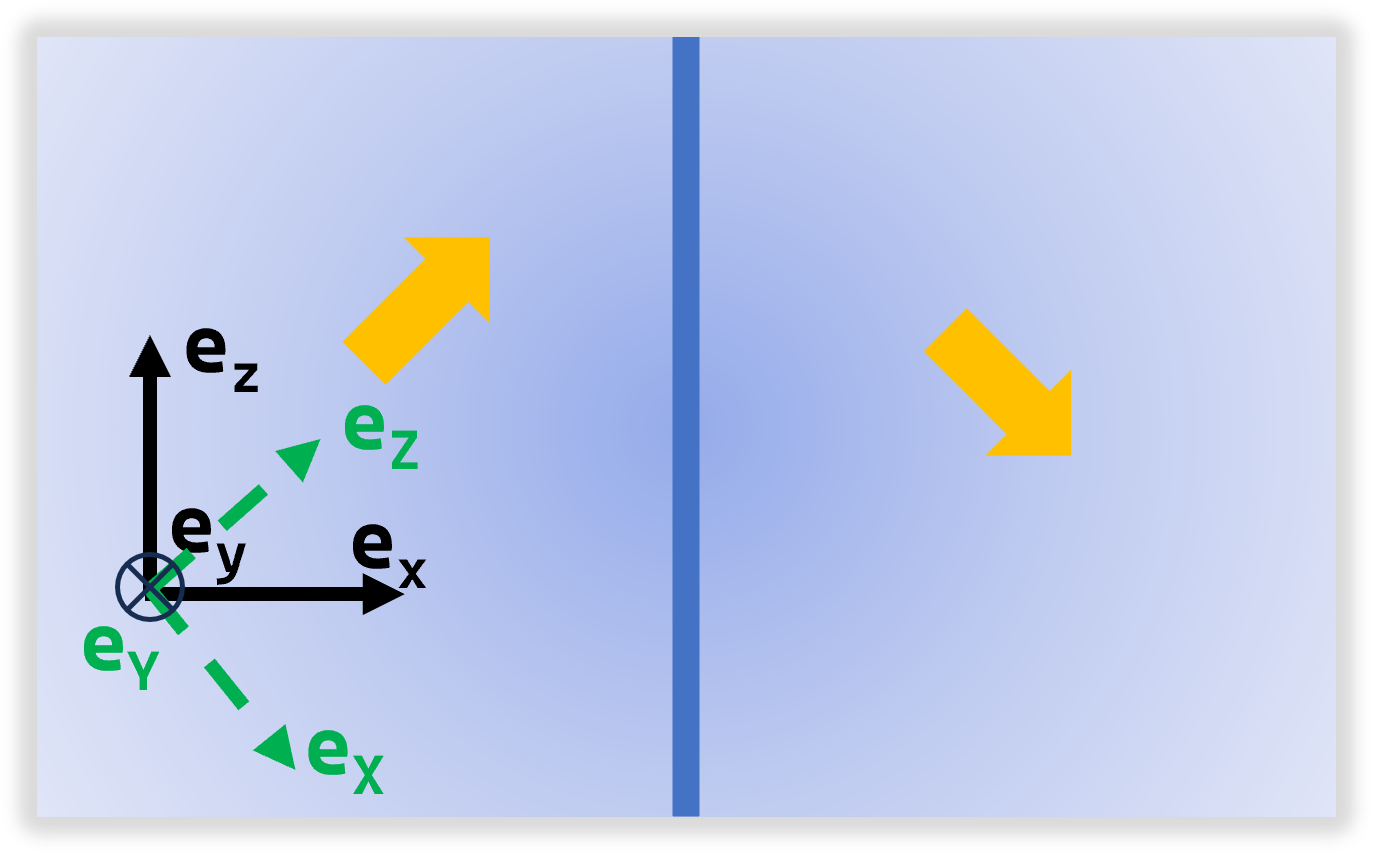}
        \caption{}
        \label{fig:90dw}
    \end{subfigure}

    \caption{Schematic cross-sectional profiles of the two ferroelectric domain walls considered in this work in the $x$--$z$ plane: (a) $180^\circ$ and (b) $90^\circ$}
    \label{fig:DWprofie}
\end{figure}

From the topology of the energy functional, which reflects the symmetries of the crystallographic unit cell of BTO, it follows that only two types of domain walls can exist: a $180^\circ$ DW, in which the polarization vectors in the two neighboring domains are antiparallel, and a $90^\circ$ DW, in which the polarization rotates by $90^\circ$ across the wall. Both types are illustrated in Fig.~\ref{fig:DWprofie} as schematic cross-sectional profiles in the $x$--$z$ plane. 

To describe the two DW profiles, we distinguish between a global coordinate system $(\mathbf e_x,\mathbf e_y,\mathbf e_z)$ and a crystal coordinate system $(\mathbf e_X,\mathbf e_Y,\mathbf e_Z)$. For the $180^\circ$ DW, the two coordinate systems coincide, so that the polarization reversal is directly represented by the variation of the $P_z$ component along the wall-normal direction $\mathbf e_x$. For the $90^\circ$ DW, the crystal frame is rotated by $45^\circ$ about the $y$-axis relative to the global frame. The polarization components in the two frames are related by
\begin{equation}
\begin{pmatrix}
P_x\\[3pt]
P_z
\end{pmatrix}
=
\mathbf R_y(\theta)
\begin{pmatrix}
P_X\\[3pt]
P_Z
\end{pmatrix},
\qquad
\mathbf R_y(\theta)=
\begin{pmatrix}
\cos\theta & \sin\theta\\[3pt]
-\sin\theta & \cos\theta
\end{pmatrix},
\qquad \theta=45^\circ.
\label{eq:rotation_P_90dw}
\end{equation}
Accordingly, the two domain states of the $90^\circ$ DW, which are naturally described in the crystal frame, are represented in the global frame by an approximately constant $P_x$ component and a sign change in $P_z$.

\subsection{Physics-Informed Neural Network}
\label{sec:pinn}

As illustrated in Section.~\ref{sec:DWs}, two independent training cases are introduced. For the $180^\circ$ DW, the wall profile is primarily governed by the $G$ and $l$, together with the Landau coefficients $a_1$, $a_2$ and $a_4$, whereas $\mu$ and $a_3$ do not play a decisive role in this switching geometry. By contrast, the $90^\circ$ DW involves polarization rotation between different crystallographic directions and is therefore sensitive to the $\mu$ and $a_3$. Therefore, the training is carried out in two stages: $(G,l,a_4)$ are identified from the $180^\circ$ DW training case, while $(\mu,a_3)$ are subsequently determined from the $90^\circ$ DW training case.

\begin{figure}[H]
  \centering
  \includegraphics[width=1\textwidth]{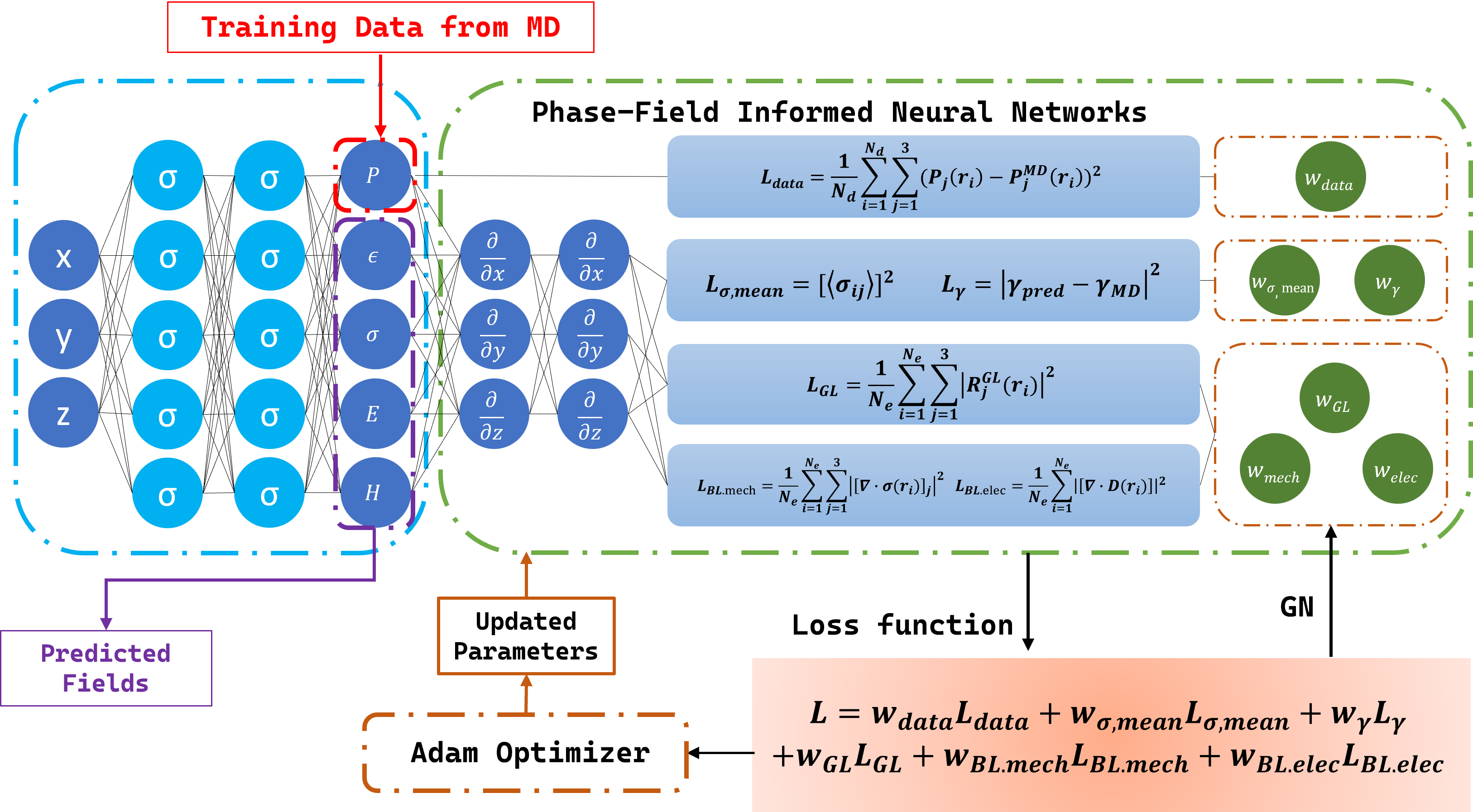}
  \caption{MD-data driven PINN training process}
  \label{fig:PINN}
\end{figure}

Fig.~\ref{fig:PINN} illustrates the workflow of the MD-data-driven Physics-Informed Neural Network (PINN). In the present work, the network is trained using stationary polarization fields associated with relaxed $180^\circ$ and $90^\circ$ domain walls (DWs) from MD, such that only equilibrium states are considered. As a result, the time-derivative term on the left-hand side of the TDGL equation in Eq.~\eqref{eq:TDGL_expanded} vanishes. The three-dimensional spatial coordinates $(x,y,z)$ are used as inputs to a fully connected residual neural network (ResNet-type multilayer perceptron, MLP), which predicts the primary solution fields in the relaxed configuration. These predicted fields are then passed to the physics module (green box in Fig.~\ref{fig:PINN}), where the governing equations and constitutive relations of the ferroelectric PFM are evaluated. Although the DW structures considered here exhibit primarily one-dimensional variation across the DW, the PINN is implemented in the full three-dimensional spatial domain using MD-derived stationary field data. However, the network outputs only the channel corresponding to the polarization component that undergoes reversal, which in the considered geometries is $P_z$ (see Figure \ref{fig:DWprofie}). 

As shown in the lower part of Fig.~\ref{fig:PINN}, training is performed by jointly optimizing both the neural network parameters (weights and biases) and selected model parameters (e.g., $a_4,G,l$) using the Adam optimizer~\cite{Kingma2015Adam}. The total loss combines supervised terms from MD data and physics-based constraints from the PFM, and gradients are obtained through automatic differentiation over the entire computational graph. In this way, the predicted fields and trainable material parameters are updated simultaneously in a closed-loop manner. The construction of the loss function will be described in detail later in this section. 

The nondimensionalization procedure used in this study is summarized in ~\ref{app:md_preprocessing}. Henceforth, all parameters, training settings, and results are presented in nondimensional form. To verify the robustness and stability of the PINN training, the initial values of the trainable parameters are generated using different random seeds. In the $180^\circ$ DW case, random initialization is applied to $G$ and $l$, whereas in the $90^\circ$ DW case, it is applied to $\mu$ and $a_3$. The corresponding sampling intervals are intentionally chosen to be far from the values reported in previous PFM studies \cite{schrade2014invariant,Durdiev2023FourierPF,schrade2007domain}, such as to test the sensitivity of the optimization to non-typical initial conditions. For $a_4$, by contrast, two representative initial values, $1.6$ and $1.9$, are selected based on the range $1.5 < a_4 < 2$ required for the topology of the phase-field energy functional to be consistent with the presence of a first-order phase transition \cite{durdiev2025parameterization}. As a result, eight representative training groups are constructed for the $180^\circ$ DW case and four for the $90^\circ$ DW case, as listed in Table~\ref{tab:pinn_init}.

\begin{table}[htbp]
\centering
\caption{Training groups defined by different nondimensional initial values}
\label{tab:pinn_init}
\setlength{\tabcolsep}{16pt}
\begin{tabular}{ccccccc}
\toprule
Case & Run & $G$ & $l$ & $a_4$ & $\mu$ & $a_3$ \\
\midrule
\multirow{6}{*}{$180^\circ$} 
& \#1 & $37.62$  & $99.19$  & $1.9$ & -- & -- \\
& \#2 & $77.27$  & $0.006$  & $1.9$ & -- & -- \\
& \#3 & $0.008$  & $58.38$  & $1.9$ & -- & -- \\
& \#4 & $0.005$  & $0.004$  & $1.9$ & -- & -- \\
& \#5 & $61.94$  & $11.46$  & $1.6$ & -- & -- \\
& \#6 & $19.81$  & $0.007$  & $1.6$ & -- & -- \\
& \#7 & $0.004$  & $29.16$  & $1.6$ & -- & -- \\
& \#8 & $0.017$  & $0.005$  & $1.6$ & -- & -- \\
\midrule
\multirow{4}{*}{$90^\circ$} 
& \#1 & -- & -- & -- & $367.73$ & $59.26$  \\
& \#2 & -- & -- & -- & $950.57$ & $0.18$ \\
& \#3 & -- & -- & -- & $0.05$   & $45.27$  \\
& \#4 & -- & -- & -- & $0.22$   & $0.05$ \\
\bottomrule
\end{tabular}
\end{table}

\paragraph{General loss function for both $180^\circ$ and $90^\circ$ DW training.}
On the right-hand side of Fig.~\ref{fig:PINN}, the physical constraints embedded in the PINN loss function are summarized. The first component, the data-fidelity term $L_{\mathrm{data}}$, penalizes the discrepancy between the predicted polarization field and the MD polarization values

\begin{equation}
L_{\mathrm{data}}
=
\frac{1}{N_{\mathrm d}}
\sum_{i=1}^{N_{\mathrm d}}
\sum_{j=1}^{3}
\left(
P_{j}(\mathbf r_i)-P^{\mathrm{MD}}_{j}(\mathbf r_i)
\right)^2 
\label{eq:L_data}
\end{equation}

Here, $\mathbf{r}_i$ indicates the spatial positions of Ti atoms in the MD simulation, and $P_{j}(\mathbf r_i)$ denotes the polarization field predicted by the PINN and interpolated at the same locations. $N_{\mathrm d}$ denotes the number of discrete polarization data points directly extracted from the MD simulations and used for supervised learning.

In this work, the polarization field is expressed in the global coordinate system, for which both the $180^\circ$ and $90^\circ$ DWs switch only in the $P_z$ component along the $x$ direction, while the change in the remaining components could be neglected. Therefore, Eq.~\ref{eq:L_data} can be simplified as

\begin{equation}
L_{\mathrm{data}}
=
\frac{1}{N_{\mathrm d}}
\sum_{i=1}^{N_{\mathrm d}}
\left(
P_{z}(\mathbf r_i)-P^{\mathrm{MD}}_{z}(\mathbf r_i)
\right)^2 .
\label{eq:L_data_simplified}
\end{equation}

We Noted that for the $90^\circ$ DW case, the $P_x$ and $E_x$ obtained from MD post-processing (in \ref{app:md_preprocessing}) is directly supplied to the network as known fields.

The second component, $L_{\mathrm{GL}}$, corresponds to the residual of the steady-state Ginzburg--Landau equation

\begin{equation}
L_{\mathrm{GL}}
=
\frac{1}{N_{\mathrm e}}
\sum_{i=1}^{N_{\mathrm e}}
\sum_{j=1}^{3}
\left|
R^{\mathrm{GL}}_{j}(\mathbf r_i)
\right|^{2}
\label{eq:L_GL_steady}
\end{equation}

\begin{equation}
\begin{aligned}
\mathbf{R}^{\mathrm{GL}}(\mathbf r_i)
&=
\frac{G}{P_s^2}
\left[\nabla \cdot \mathbb{L} : (\nabla \mathbf{P})\right]_{\mathbf r=\mathbf r_i}
-
G \left.\frac{\partial \psi(\mathbf{P})}{\partial \mathbf{P}}\right|_{\mathbf r=\mathbf r_i}
-
\left.\frac{\partial H^{\mathrm{bulk}}}{\partial \mathbf{P}}\right|_{\mathbf r=\mathbf r_i}
\end{aligned}
\label{eq:GL_residual_steady}
\end{equation}

where $\mathbf{R}^{\mathrm{GL}}(\mathbf r_i) = \bigl(R^{\mathrm{GL}}_{1}(\mathbf r_i), R^{\mathrm{GL}}_{2}(\mathbf r_i), R^{\mathrm{GL}}_{3}(\mathbf r_i)\bigr)$
denotes the vector-valued residual of the steady-state Ginzburg--Landau equation evaluated at the collocation point $\mathbf r_i$, with $R^{\mathrm{GL}}_{j}$ corresponding to the $j$-th Cartesian component of the polarization field. $N_{\mathrm e}$ denotes the number of equation points, which is chosen as $N_{\mathrm e} = 5\,N_{\mathrm d}$ in order to impose the physical constraints more densely in the spatial domain than the available MD data.

The third and the fourth term, $L_{\mathrm{BL.mech}}$ and $L_{\mathrm{BL.elec}}$, enforce mechanical and electrostatic balance by penalizing the residuals of the corresponding equilibrium equations, respectively
\begin{equation}
L_{\mathrm{BL.mech}}
=
\frac{1}{N_{\mathrm e}}
\sum_{i=1}^{N_{\mathrm e}}
\sum_{j=1}^{3}
\left|
\left[\nabla \cdot \boldsymbol{\sigma}(\mathbf r_i)\right]_j
\right|^{2},
\qquad
L_{\mathrm{BL.elec}}
=
\frac{1}{N_{\mathrm e}}
\sum_{i=1}^{N_{\mathrm e}}
\left|
\nabla \cdot \mathbf{D}(\mathbf r_i)
\right|^{2}.
\label{eq:L_BL}
\end{equation}

In addition, to ensure physical consistency with the MD--NPT ensemble, we introduce a penalty term that enforces vanishing mean stress
\begin{equation}
L_{\boldsymbol{\sigma}_{\mathrm{mean}}}
=
\left\|\left\langle \boldsymbol{\sigma}^{\mathrm{pred}} \right\rangle\right\|_2^2
\end{equation}
where $\langle \cdot \rangle$ denotes the spatial average over the computational domain.

To further constrain the DW energetics, we introduce an additional loss term
\begin{equation}
L_{\gamma}
=
\left(\gamma^{\mathrm{pred}}-\gamma^{\mathrm{ref}}\right)^2,
\label{eq:loss_gamma}
\end{equation}
where $\gamma^{\mathrm{ref}}$ denotes the reference DW energy obtained from MD (see Table~\ref{tab:parameters}) for the $180^\circ$ and $90^\circ$ DW cases, and $\gamma^{\mathrm{pred}}$ is computed from the PINN-predicted fields as~\cite{Dieguez2022TranslationalCovarianceFlexoelectricity}
\begin{equation}
\gamma^{\mathrm{pred}}
=
\int_{x_0}^{x_1}
\left[
H(x)-H_{\mathrm{ref}}
\right]\mathrm{d}x
\label{eq:dw_energy_1d}
\end{equation}
Here, $x$ is the coordinate along the wall-normal direction, and $x_0$ and $x_1$ denote the endpoints of the one-dimensional integration window. In practice, the interval $[x_0,x_1]$ is chosen such that it fully covers the DW core and extends sufficiently into the far-field regions on both sides, where $H(x)\approx H_{\mathrm{ref}}$, i.e., the saturated bulk domains.

The total loss is defined as
\begin{equation}
\begin{aligned}
L_{\mathrm{total}}
&=
w_{\mathrm{data}} L_{\mathrm{data}}
+
w_{\boldsymbol{\sigma}_{\mathrm{mean}}} L_{\boldsymbol{\sigma}_{\mathrm{mean}}}
+
w_{\gamma} L_{\gamma}
\\
&\quad
+
w_{\mathrm{GL}} L_{\mathrm{GL}}
+
w_{\mathrm{BL.elec}} L_{\mathrm{BL.elec}}
+
w_{\mathrm{BL.mech}} L_{\mathrm{BL.mech}}
\end{aligned}
\label{eq:total_loss}
\end{equation}
Here, $w_i$ (with $i \in \{\mathrm{data}, \boldsymbol{\sigma}_{\mathrm{mean}}, \gamma, \mathrm{GL}, \mathrm{BL.elec}, \mathrm{BL.mech}\}$) denote the weighting coefficients associated with the corresponding loss terms. In the present implementation, these coefficients are divided into three categories.

First, the data weight $w_{\mathrm{data}}$ is adjusted dynamically during training by a self-designed strategy. This treatment is introduced because repeated numerical tests showed that the data loss $L_{\mathrm{data}}$ is highly sensitive to the initial values of $G$ and $l$. In particular, when the initial values of $G$ and $l$ are relatively large, the network tends to have difficulty recovering a physically reasonable DW proflie. Since preserving the DW shape is essential for stabilizing the subsequent parameter identification process, the data term is strengthened adaptively whenever the supervised polarization fitting becomes insufficient. Specifically, $L_{\mathrm{data}}$ is monitored during training, and if
$L_{\mathrm{data}} > 10^{-4}$ persists for 50 consecutive monitoring steps, $w_{\mathrm{data}}$ is increased linearly according to
\begin{equation}
w_{\mathrm{data}}^{(n+1)}
=
w_{\mathrm{data}}^{(n)}
+
\Delta w_{\mathrm{data}}^{+}
\label{eq:wdata_update}
\end{equation}
where $\Delta w_{\mathrm{data}}^{+}$ is a prescribed positive increment and $n$ denotes the update index. In this way, the supervised data constraint is reinforced only when needed, which helps maintain the correct DW morphology while reducing undesired interference from unfavorable initial parameter values.

To prevent $w_{\mathrm{data}}$ from becoming excessively large and causing the network to neglect other physics-related loss terms, an additional relaxation criterion is introduced. If
$L_{\mathrm{data}} < 10^{-5}$ persists for 100 consecutive monitoring steps, the data weight is reduced linearly as
\begin{equation}
w_{\mathrm{data}}^{(n+1)}
=
w_{\mathrm{data}}^{(n)}
-
\Delta w_{\mathrm{data}}^{-}
\label{eq:wdata_decrease}
\end{equation}
where $\Delta w_{\mathrm{data}}^{-}$ is a prescribed positive decrement.

Second, the weights $w_{\boldsymbol{\sigma}_{\mathrm{mean}}}$ and $w_{\gamma}$ are fixed manually before training. In practice, their values are selected empirically by trial-and-error to ensure that both constraints remain effective throughout training.

Third, the weights $w_{\mathrm{GL}}$, $w_{\mathrm{BL.elec}}$, and $w_{\mathrm{BL.mech}}$ are updated adaptively during training using the GradNorm method. These three terms correspond to the governing-equation residuals of the PFM, whose relative scales and convergence rates can differ substantially. Therefore, rather than prescribing their weights a priori, they are adjusted automatically by GradNorm~\cite{chen2018gradnorm} so that the three physics-based residual losses evolve at comparable training rates. The detailed formulation and implementation of the GradNorm procedure used in this work are provided in~\ref{app:gradnorm}.

\paragraph{Case-specific treatment for $180^\circ$ DW training}

For the $180^\circ$ DW training case, besides the general loss constraints described above, the coefficient relations among $a_1$, $a_2$, and $a_4$ introduced in the Section ~\ref{sec:pfm} (Eq.[\ref{eq:a124}]) are enforced. Therefore, only $a_4$ needs to be learned, while $a_1$ and $a_2$ are derived from these relations, which reduce the number of independent degrees of freedom during training.

To improve the robustness and identifiability of parameter training in the $180^\circ$DW case, we split the optimization procedure into three stages instead of updating all trainable quantities simultaneously from the beginning, which is inspired by the idea of curriculum learning and continuation-type strategies reported in \cite{Huang2022HomPINNs, Krishnapriyan2021FailureModesPINN, Dwivedi2025PIELM}. This staged strategy is motivated by two practical issues observed in the present problem. First, $G$ and $l$ are strongly coupled in the governing equations (and in the effective gradient-energy contribution), which makes their simultaneous optimization prone to compensation effects. In practice, different combinations of $G$ and $l$ may produce similar wall profiles during early training, resulting in unstable parameter drift and poor reproducibility. Second, under the MD-consistent NPT setting adopted here, the elastic-energy contribution is relatively weak compared with the dominant polarization-related terms. As a result, if all parameters are released from the start, the optimizer may preferentially reduce the data and polarization-related residuals while driving $(G,l)$ along an undesirable coupled path, which can lead to an unrealistic wall width or distorted parameter evolution. The details of the stepwise training are  given in ~\ref{app:3step}.

\paragraph{Case-specific treatment for $90^\circ$ DW training.}
For the $90^\circ$ DW case, we do not infer the electrostatic field from the network, which could decrease the freedom of degree of the training. Instead, the MD-derived electric field $\mathbf{E}^{\mathrm{MD}}(\mathbf{r})$ is imposed as a known input and is directly used in the PINN (see details of the preprocessing in ~\ref{app:md_preprocessing}). 

To further regularize the separation-energy landscape along the $90^\circ$ switching pathway, we introduce an additional constraint associated with the transition state connecting the two polarization minima. Physically, this transition state represents the critical intermediate configuration that the system must pass through during polarization switching. Along the minimum-energy switching path, it corresponds to the highest-energy point separating the two stable polarization states. From the viewpoint of the multidimensional energy landscape, this transition state is the saddle point between the two polarization minima. The corresponding polarization at this saddle point, in nondimensional form, is taken as
\begin{equation}
\mathbf{P}^{\mathrm{ts}} = (P^{\mathrm{ts}}_x,P^{\mathrm{ts}}_y,P^{\mathrm{ts}}_z)= P_s(0.528,0,0.528),
\end{equation}
as determined from the approximate $90^\circ$ switching path extracted from the MD data (see \ref{app:90dw_switching_path}).
A stationary-point condition is imposed by penalizing the first derivatives of the separation energy density $H_{\mathrm{sep}}$ evaluated at $\mathbf{P}^{\mathrm{ts}}$:
\begin{equation}
L_{\mathrm{stat}}
=
\left(
\left.\frac{\partial H_{\mathrm{sep}}}{\partial P_x}\right|_{\mathbf{P}=\mathbf{P}^{\mathrm{ts}}}
\right)^2
+
\left(
\left.\frac{\partial H_{\mathrm{sep}}}{\partial P_z}\right|_{\mathbf{P}=\mathbf{P}^{\mathrm{ts}}}
\right)^2
\label{eq:loss_stat_90dw}
\end{equation}

Since a vanishing first derivative alone only guarantees a stationary point, we further characterize the local energy-surface geometry through the Hessian matrix of $H_{\mathrm{sep}}$ in the $(P_x,P_z)$ plane:
\begin{equation}
\mathbf{H}_{\mathrm{sep}}
=
\begin{pmatrix}
\displaystyle
\left.\frac{\partial^2 H_{\mathrm{sep}}}{\partial P_x^2}\right|_{\mathbf{P}=\mathbf{P}^{\mathrm{ts}}}
&
\displaystyle
\left.\frac{\partial^2 H_{\mathrm{sep}}}{\partial P_x \partial P_z}\right|_{\mathbf{P}=\mathbf{P}^{\mathrm{ts}}}
\\[6pt]
\displaystyle
\left.\frac{\partial^2 H_{\mathrm{sep}}}{\partial P_z \partial P_x}\right|_{\mathbf{P}=\mathbf{P}^{\mathrm{ts}}}
&
\displaystyle
\left.\frac{\partial^2 H_{\mathrm{sep}}}{\partial P_z^2}\right|_{\mathbf{P}=\mathbf{P}^{\mathrm{ts}}}
\end{pmatrix}
\label{eq:hessian_sep_90dw}
\end{equation}

To impose a saddle-like local structure, we project the Hessian onto the approximate switching direction $\mathbf e_{\parallel}$ and its orthogonal direction $\mathbf e_{\perp}$ in the $(P_x,P_z)$ plane:
\begin{equation}
\lambda_{\parallel}
=
\mathbf e_{\parallel}^{\mathrm T}\mathbf H_{\mathrm{sep}}\mathbf e_{\parallel},
\qquad
\lambda_{\perp}
=
\mathbf e_{\perp}^{\mathrm T}\mathbf H_{\mathrm{sep}}\mathbf e_{\perp}
\label{eq:lambda_para_perp_90dw}
\end{equation}
For the present implementation, $\mathbf e_{\parallel}$ is taken as the approximate $90^\circ$ switching-path direction in the $(P_x,P_z)$ plane, and $\mathbf e_{\perp}$ is chosen as the corresponding orthogonal direction. Negative curvature along $\mathbf e_{\parallel}$ and positive curvature along $\mathbf e_{\perp}$ are enforced through
\begin{equation}
L_{\parallel}
=
\left[\max\!\left(\lambda_{\parallel}+\delta_{-},\,0\right)\right]^2,
\qquad
L_{\perp}
=
\left[\max\!\left(\delta_{+}-\lambda_{\perp},\,0\right)\right]^2
\label{eq:loss_para_perp_90dw}
\end{equation}
where $\delta_{-}>0$ and $\delta_{+}>0$ are small margin parameters.

In addition, a weak determinant-based constraint is introduced to exclude locally convex configurations:
\begin{equation}
L_{\det}
=
\left[\max\!\left(\det(\mathbf H_{\mathrm{sep}})+\delta_{\det},\,0\right)\right]^2
\label{eq:loss_det_90dw}
\end{equation}
where $\delta_{\det}>0$ is a small margin.

The final transition-state regularization is then written as
\begin{equation}
L_{\mathrm{saddle}}
=
L_{\mathrm{stat}}
+
L_{\det}
+
L_{\parallel}
+
L_{\perp}
\label{eq:loss_saddle_90dw}
\end{equation}
Similar to the $180^\circ$ DW training case, we adopt a staged optimization strategy rather than releasing all trainable quantities simultaneously. This choice is motivated by the strong nonlinear coupling between the  Landau coefficient $a_3$ and the gradient energy anisotropy parameter $\mu$. Specifically, $a_3$ controls the barrier height of the separation-energy landscape (Eq.~\ref{eq:landauPoly}) near the saddle point along the $90^\circ$ polarization switching path. A larger $a_3$ tends to increase the separation energy penalty associated with the switching region and therefore favors a sharper DW profile. To maintain a physically reasonable DW width consistent with the MD reference, this tendency must be compensated by a sufficiently strong gradient regularization, governed here by the gradient-energy tensor Eq~\ref{eq:anisotropic_tensor}. As a result, $a_3$ and $\mu$ become strongly coupled during optimization. If both parameters are released simultaneously from the beginning, the optimizer often encounters an ill-conditioned loss landscape, making it difficult to balance the localized barrier-height constraint and the global gradient regularization in a stable manner. The details of the stepwise training protocol are found in \ref{app:3step}.

\subsection{Finite Element Implementation}
\label{sec:fem}

To independently validate the PFM parameters inferred by the PINN, we solve the ferroelectric PFM using the finite element method (FEM) as a reference. The primary field variables are the spontaneous polarization $\mathbf{P}$, the mechanical displacement $\mathbf{u}$, and the electric potential $\phi$, governed by the coupled PDEs summarized in Section~\ref{sec:pfm}.

\begin{figure}[H]
  \centering
  \includegraphics[width=0.75\textwidth]{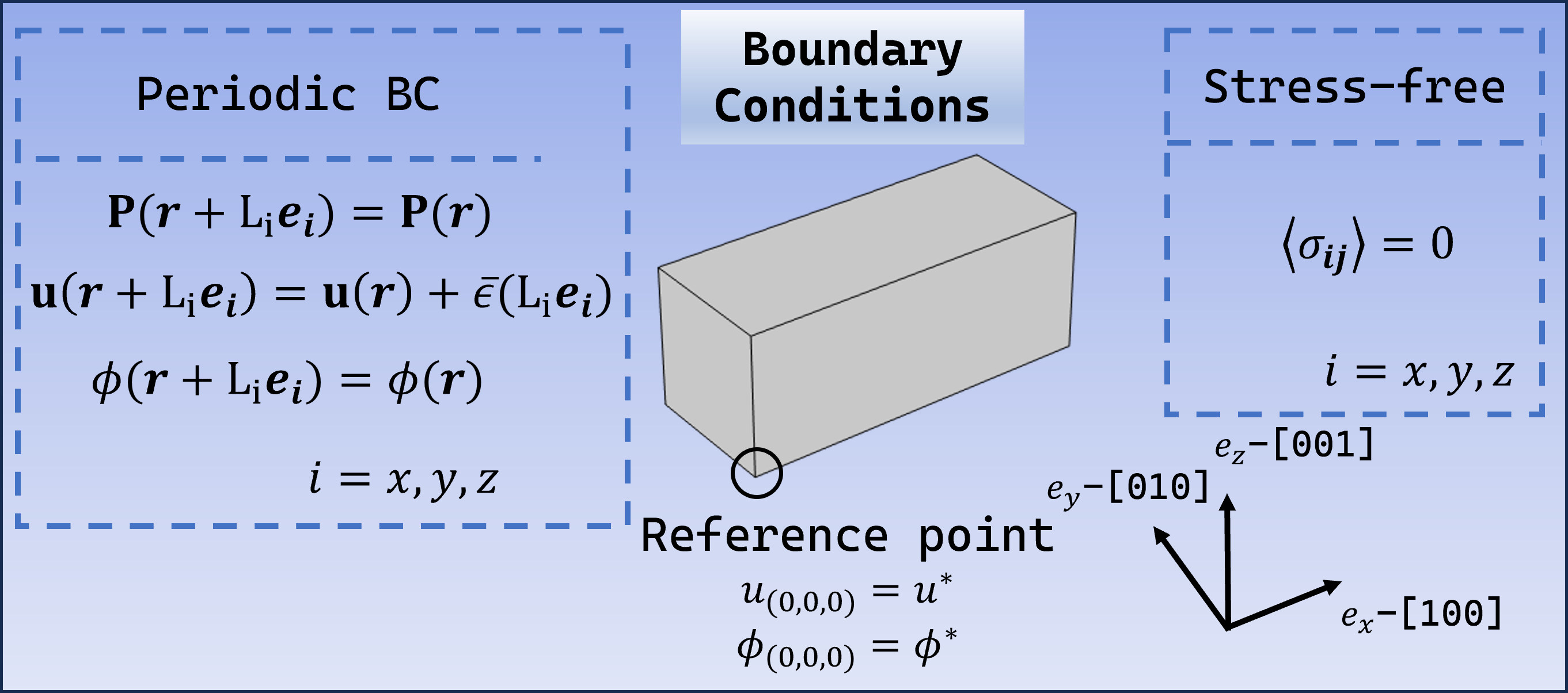}
  \caption{Simulation domain and boundary condition setup used in FEM.}
  \label{fig:BCs}
\end{figure}

Fig~\ref{fig:BCs} illustrates the simulation domain and the coordinate system, where $x$, $y$, and $z$ correspond to the crystallographic directions [100], [010], and [001], respectively. The FEM domain matches the MD supercell dimensions, with $L_y=L_z=4.0\,\mathrm{nm}$, while $L_x=10\,\mathrm{nm}$ for the $180^\circ$ DW case and $L_x=20\,\mathrm{nm}$ for the $90^\circ$ DW case. All field variables are discretized using cubic Lagrange shape functions on three-dimensional hexahedral elements. The grid spacing is set to $0.2\,\mathrm{nm}$, resulting in a total of $20{,}000$ and $40{,}000$ elements for $180^\circ$ and $90^\circ$ simulation cases, respectively.

To be consistent with the atomistic setup, we apply the following boundary conditions: (i) a reference-point constraint is introduced to remove rigid-body drift and fix the gauge freedom of the displacement and electric potential fields; and (ii) periodic boundary conditions are enforced along all three spatial directions for $\mathbf{P}$, $\phi$, and the periodic part of the displacement field $\mathbf{u}^{\mathrm{per}}$. 

To reproduce the stress-free mechanical environment within a fully periodic cell, the displacement field is decomposed as
\begin{equation}
\mathbf{u}(\mathbf{r})=\mathbf{u}^{\mathrm{per}}(\mathbf{r})+\bar{\boldsymbol{\varepsilon}}\,\mathbf{r},
\label{eq:u_decomposition}
\end{equation}
where $\mathbf{u}^{\mathrm{per}}$ is periodic and $\bar{\boldsymbol{\varepsilon}}$ is the macroscopic strain tensor. The components of $\bar{\boldsymbol{\varepsilon}}$ are treated as global unknowns and determined by enforcing vanishing volume-averaged normal stresses
\begin{equation}
\langle\sigma_{xx}\rangle=0,\qquad
\langle\sigma_{yy}\rangle=0,\qquad
\langle\sigma_{zz}\rangle=0,
\label{eq:mean_stress_zero}
\end{equation}
thereby allowing the periodic simulation box to relax its shape while maintaining periodicity.

Time-dependent FEM simulations are performed starting from an initially sharp DW profile until a stationary configuration is reached. Here, the stationary state is numerically defined by the convergence of the Backward Differentiation Formula (BDF) solver in COMSOL Multiphysics 6.3, where the relative error of all dependent variables falls below a strict tolerance of $10^{-3}$. The converged FEM results are then compared with the time-averaged MD polarization fields obtained over the converged portion of the trajectories.

\section{Results and Discussion}
\label{sec:rd}

\subsection{Parameter Identification}

\begin{figure}[H]
    \centering

    \begin{subfigure}[b]{0.4\textwidth}
        \centering
        \includegraphics[width=\linewidth]{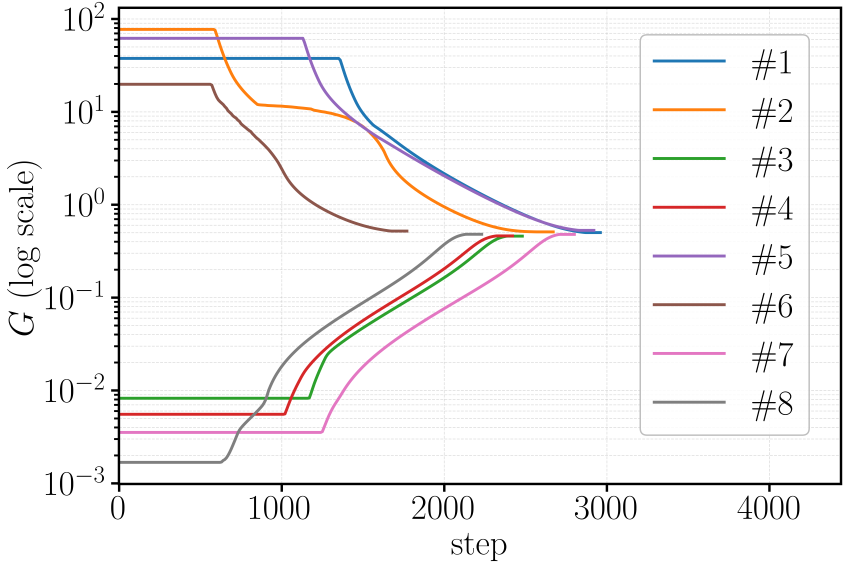}
        \caption{}
    \end{subfigure}
    \hspace{0.01\textwidth}
    \begin{subfigure}[b]{0.4\textwidth}
        \centering
        \includegraphics[width=\linewidth]{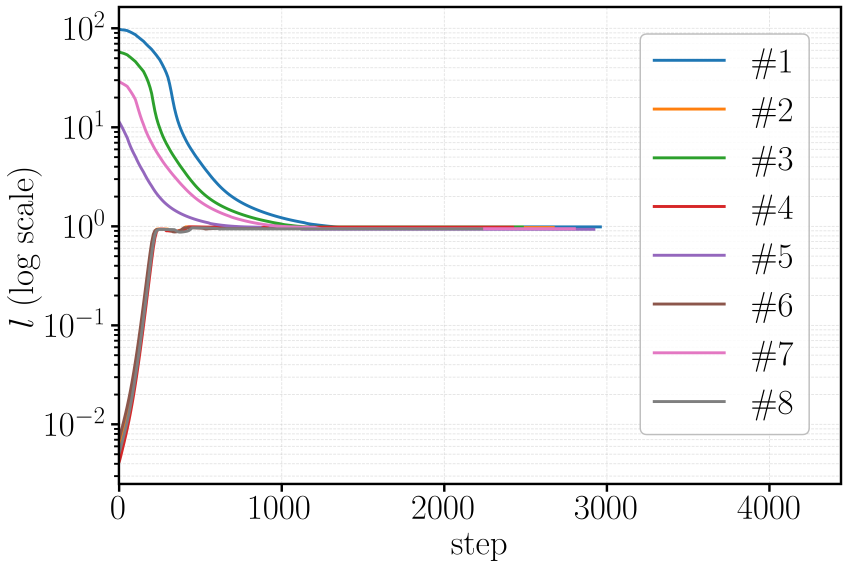}
        \caption{}
    \end{subfigure}
    \hspace{0.01\textwidth}
    \begin{subfigure}[b]{0.4\textwidth}
        \centering
        \includegraphics[width=\linewidth]{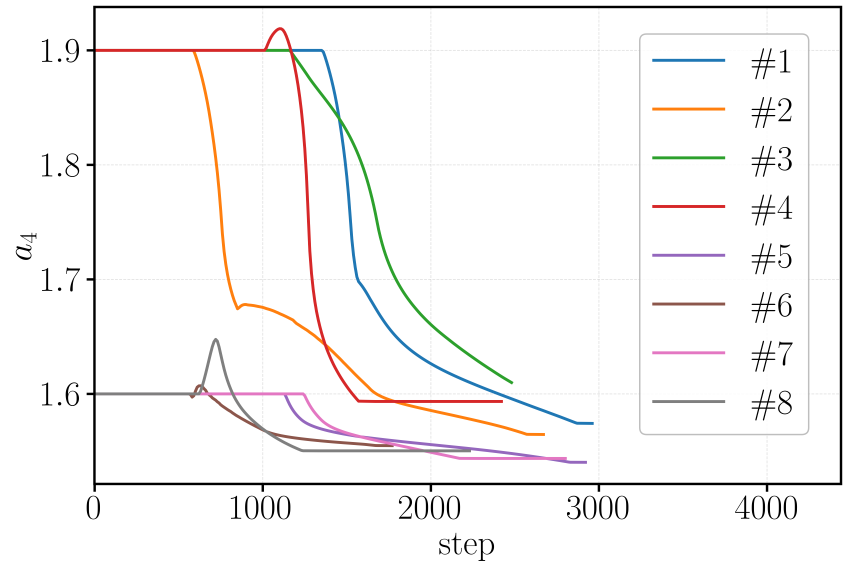}
        \caption{}
    \end{subfigure}
    \begin{subfigure}[b]{0.4\textwidth}
        \centering
        \includegraphics[width=\linewidth]{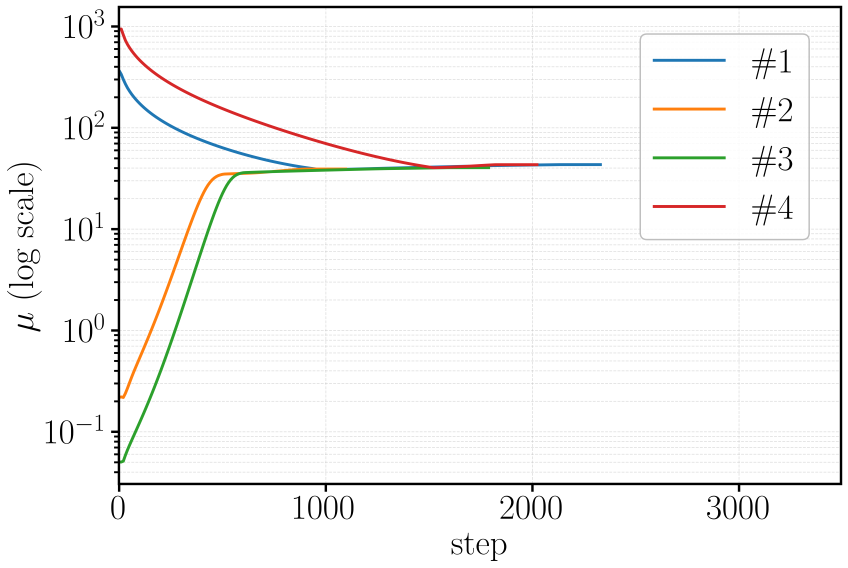}
        \caption{}
    \end{subfigure}
    \hspace{0.01\textwidth}
    \begin{subfigure}[b]{0.4\textwidth}
        \centering
        \includegraphics[width=\linewidth]{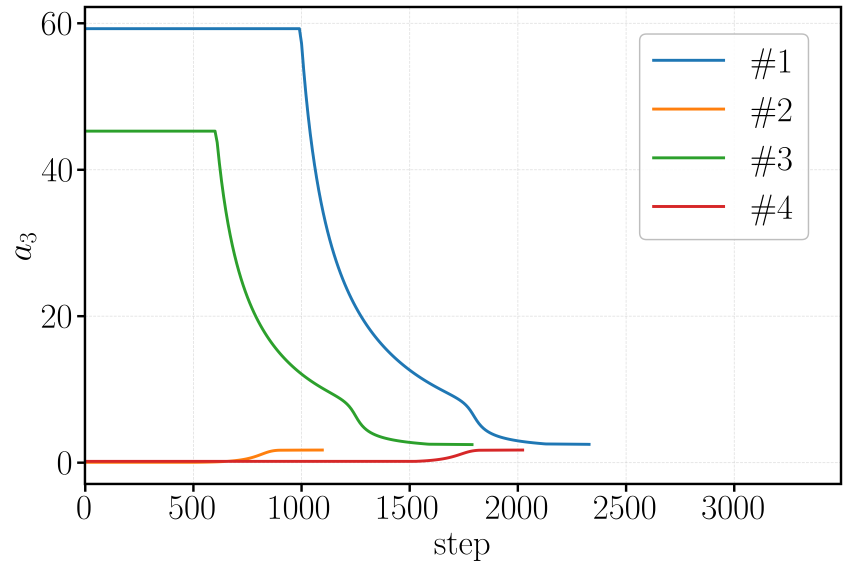}
        \caption{}
    \end{subfigure}

    \caption{Parameters evolution of each parameter during $180^\circ$ and $90^\circ$ DWs training: 
    (a) $G$
    (b) $l$ and
    (c) \(a_4\)
    (d) $\mu$
    (e) \(a_3\)}
    
    \label{fig:IDParameters}
\end{figure}

Figure~\ref{fig:IDParameters} summarizes the parameter evolution during the staged training procedure. Despite the large variation in the initial guesses, all parameters progressively approach a narrow range of terminal values, indicating that the proposed multi-stage strategy effectively reduces the sensitivity to initialization. For the \(180^\circ\) DW-related parameters, \(l\) converges most rapidly, which is consistent with the first training stage, where the simplified GL residual primarily anchors the wall width and profile. By contrast, \(G\) and \(a_4\) are only fully activated after the first stage, when the complete loss function is introduced. Once activated, \(G\) evolves monotonically, either increasing or decreasing depending on the initialization, and finally approaches a stable value. The evolution of \(a_4\) is less direct: in several runs it exhibits a brief initial increase, followed by a gradual decrease towards its terminal value. For the \(90^\circ\) DW-related parameters, \(\mu\) exhibits a clear monotonic convergence towards a common terminal range across all runs. Similarly, after \(a_3\) is activated in the second stage, it evolves monotonically, either increasing or decreasing depending on the initialization, and gradually converges towards a similar terminal range across different runs.

Figure~\ref{fig:IDParameters} summarizes the parameter evolution during the staged training procedure. Despite the large variation in the initial guesses, all parameters progressively approach a narrow range of terminal values, indicating that the proposed multi-stage strategy effectively reduces the sensitivity to initialization. For the \(180^\circ\) DW-related parameters, \(l\) converges most rapidly, which is consistent with the first training stage, where the simplified GL residual primarily anchors the wall width and profile. By contrast, \(G\) and \(a_4\) are only fully activated after the first stage, when the complete loss function is introduced. Once activated, \(G\) evolves monotonically, either increasing or decreasing depending on the initialization, and finally approaches a stable value. The evolution of \(a_4\) is less direct: in several runs it exhibits a brief initial increase, followed by a gradual decrease towards its terminal value. 

To further evaluate the variability of the identified parameters across different runs, the relative range (RR), standard deviation (SD), and coefficient of variation (CV) are calculated. They are defined as
\begin{equation}
RR = \frac{q_{\max} - q_{\min}}{\bar{q}} \times 100\%,
\end{equation}
\begin{equation}
SD = \sqrt{\frac{1}{n-1}\sum_{i=1}^{n}(q_i-\bar{q})^2},
\end{equation}
and
\begin{equation}
CV = \frac{SD}{\bar{q}} \times 100\%,
\end{equation}
where \(q_i\) is the identified value from the \(i\)-th run, \(q_{\max}=\max(q_i)\), \(q_{\min}=\min(q_i)\), \(\bar{q}\) is the mean value, and \(n\) is the number of runs.

Table.~\ref{tab:pinn_param_result} provides a quantitative assessment of the run-to-run variability and thereby supports the convergence behavior observed in Fig.~\ref{fig:IDParameters}. For the \(180^\circ\) DW-related parameters, all three indicators remain low, with CV values below \(2\%\), showing that \(G\), \(l\), and \(a_4\) are identified in a highly reproducible manner despite the large spread in the initial guesses. Among them, \(G\) and \(l\) exhibit particularly small relative scatter, while \(a_4\) shows only slightly larger CV, consistent with its somewhat more involved and occasionally non-monotonic evolution during training. For the \(90^\circ\) DW-related parameters, the variability is also moderate: although \(\mu\) displays a larger spread than the \(180^\circ\) DW parameters, its CV still remains at the level of only a few percent, while \(a_3\) is recovered with comparably low relative variation. Overall, Table.~\ref{tab:pinn_param_result} quantitatively confirms that the proposed staged training strategy yields robust and reproducible parameter estimates across independent runs, and thus provides statistical support for the qualitative convergence trends shown in Fig.~\ref{fig:IDParameters}. In summary, the mean values of the trained parameters in physical units are $G = 6.0 \times 10^6~\mathrm{J/m^3}$, $l = 0.387~\mathrm{nm}$, $a_4 = 1.566$, $\mu = 41.53$, and $a_3 = 1.74$, which are consistent with the stable convergence behavior discussed above. These mean values are subsequently used in the FEM simulations.

\begin{table}[htbp]
\centering
\caption{Identificated Parameters (nondimensionlized) from PINN}
\label{tab:pinn_param_result}
\setlength{\tabcolsep}{16pt}
\begin{tabular}{ccccccc}
\toprule
Case & Run & $G$ & $l$ & $a_4$ & $\mu$ & $a_3$ \\
\midrule
\multirow{12}{*}{$180^\circ$} 
& \#1 & $0.501$  & $0.987$   & $1.574$ & -- & -- \\
& \#2 & $0.510$  & $0.972$   & $1.565$ & -- & -- \\
& \#3 & $0.499$  & $0.987$   & $1.610$ & -- & -- \\
& \#4 & $0.489$  & $0.981$   & $1.593$ & -- & -- \\
& \#5 & $0.516$  & $0.944$   & $1.540$ & -- & -- \\
& \#6 & $0.499$  & $0.954$   & $1.555$ & -- & -- \\
& \#7 & $0.509$  & $0.955$   & $1.544$ & -- & -- \\
& \#8 & $0.498$  & $0.958$   & $1.550$ & -- & -- \\
\cdashline{2-7}
& Mean    & $0.503$  & $0.967$   & $1.566$ & -- & -- \\
& RR (\%) & $5.37$   & $4.45$    & $4.47$  & -- & -- \\
& SD      & $0.009$  & $0.017$   & $0.025$ & -- & -- \\
& CV (\%) & $1.77$   & $1.76$    & $1.60$  & -- & -- \\
\midrule
\multirow{8}{*}{$90^\circ$} 
& \#1 & -- & -- & -- & $43.31$ & $1.78$ \\
& \#2 & -- & -- & -- & $39.17$ & $1.72$ \\
& \#3 & -- & -- & -- & $40.4$ & $1.76$ \\
& \#4 & -- & -- & -- & $43.25$ & $01.71$ \\
\cdashline{2-7}
& Mean    & -- & -- & -- & $41.53$ & $1.74$ \\
& RR (\%) & -- & -- & -- & $9.97$  & $4.02$ \\
& SD      & -- & -- & -- & $2.09$  & $0.03$ \\
& CV (\%) & -- & -- & -- & $5.03$  & $1.89$ \\
\bottomrule
\end{tabular}
\end{table}

\subsection{\texorpdfstring{Reconstruction of $180^\circ$ Domain Wall Fields}{Reconstruction Fields of $180^\circ$ Domain Wall}}

\label{sec:180degreeDWs}

\begin{figure}[H]
    \centering
    \includegraphics[width=0.45\textwidth]{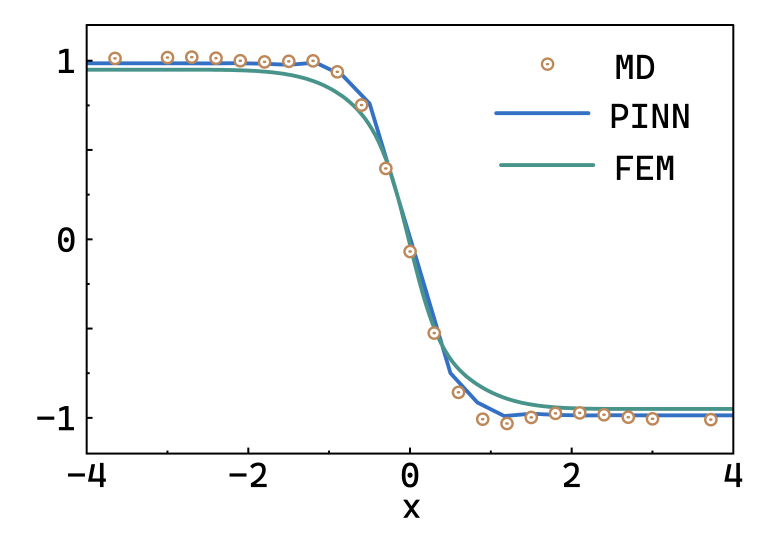}
    \caption{One-dimensional polarization profile of the $180^\circ$ domain wall from MD, PINN, and FEM.}
    \label{fig:180dw_polarization}
\end{figure}

Fig.~\ref{fig:180dw_polarization} compares the one-dimensional polarization profiles obtained from MD, PINN, and FEM along the central line \( y = z = 0.5L_x = 0.5L_y \). The PINN prediction is in near-perfect agreement with the MD profile, which is also consistent with the convergence behavior observed in the loss curves. By contrast, the FEM curve is smoother, reflecting the intrinsic characteristics of the continuum formulation, while still capturing the essential features of the \(180^\circ\) DW.

\begin{figure}[H]
    \centering

    \begin{subfigure}[b]{0.45\textwidth}
        \centering
        \includegraphics[width=\linewidth]{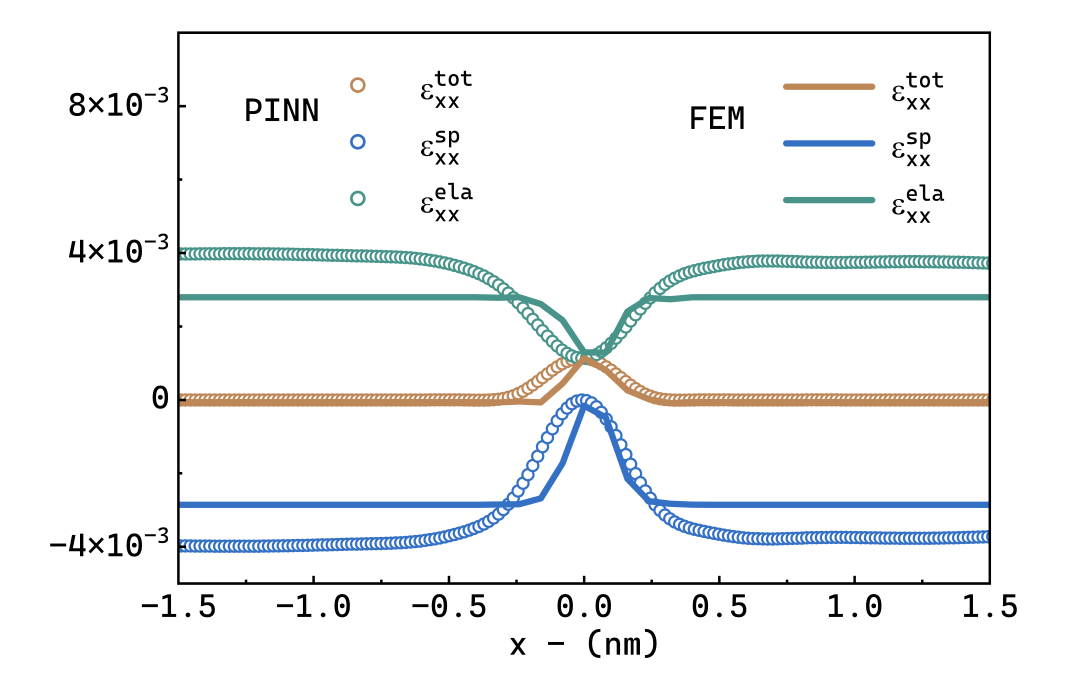}
        \caption{}
        \label{fig:180dwepsxx}
    \end{subfigure}
    \hspace{0.01\textwidth}
    \begin{subfigure}[b]{0.45\textwidth}
        \centering
        \includegraphics[width=\linewidth]{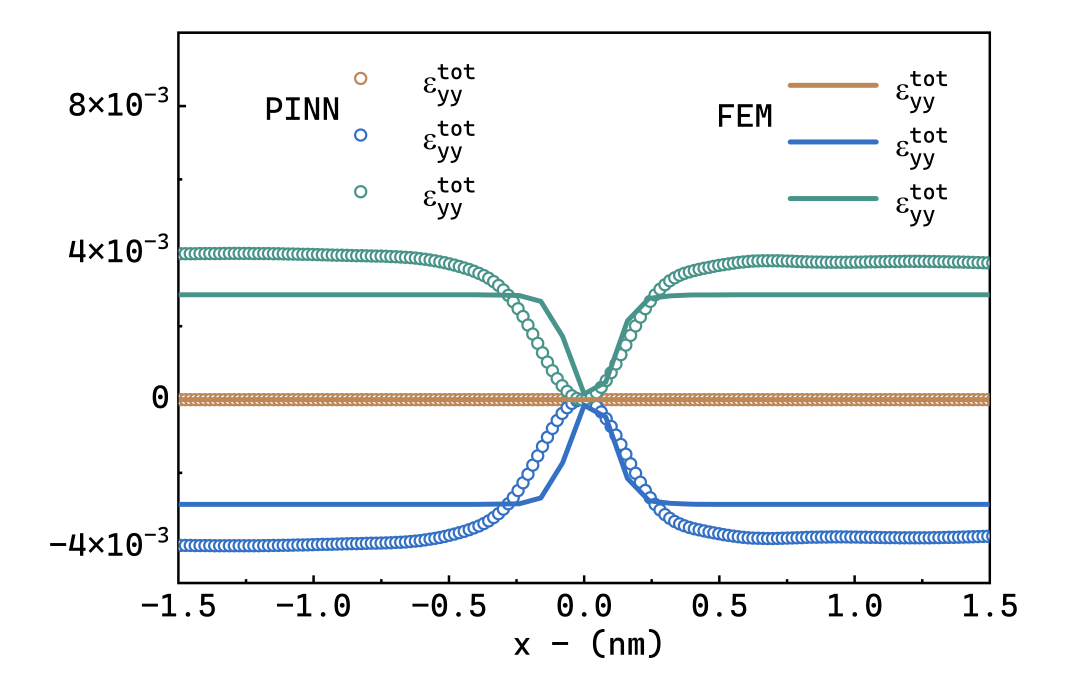}
        \caption{}
        \label{fig:180dwepsyy}
    \end{subfigure}
    \hspace{0.01\textwidth}
    \begin{subfigure}[b]{0.45\textwidth}
        \centering
        \includegraphics[width=\linewidth]{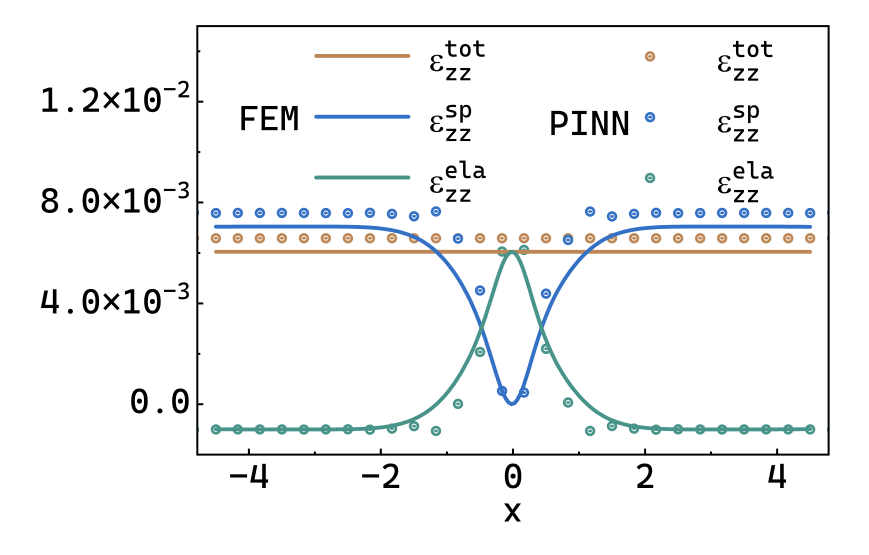}
        \caption{}
        \label{fig:180dwepszz}
    \end{subfigure}
    \hspace{0.01\textwidth}
    \begin{subfigure}[b]{0.43\textwidth}
        \centering
        \includegraphics[width=\linewidth]{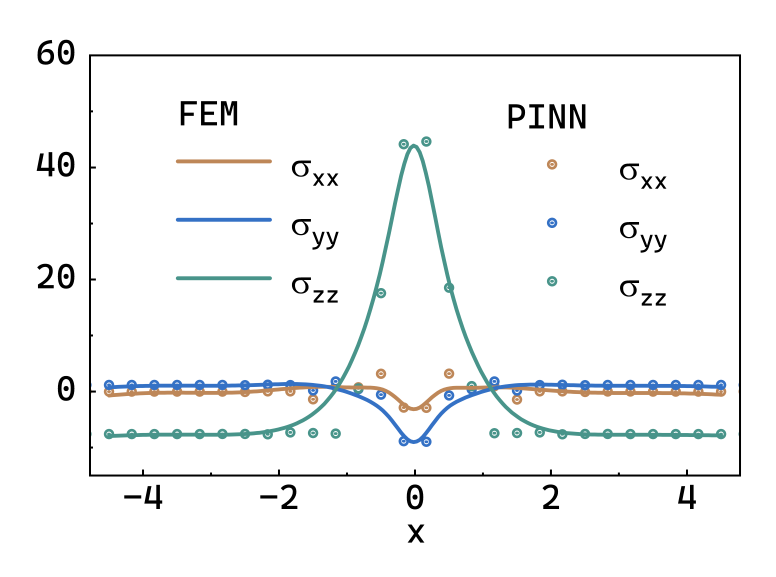}
        \caption{}
        \label{fig:180dwstress}
    \end{subfigure}

    \caption{Strain/stress fields output from both PINN and FEM of $180^\circ$ DW case: 
(a) $\varepsilon_{xx}$
(b) $\varepsilon_{yy}$
(c) $\varepsilon_{zz}$
(d) $\sigma_{xx}$, $\sigma_{yy}$, and $\sigma_{zz}$}
    \label{fig:PINNStrainStressField}
\end{figure}

As shown in Fig.~\ref{fig:PINNStrainStressField}, the strain/stress distributions predicted by the PINN are in good agreement with the FEM reference for the $180^\circ$ DW case. These results are consistent with the expected behavior of the MD-NPT simulations which were conducted with periodic boundary conditidons and at zero average stress. In the saturated polarization regions far from the domain wall, the material exhibits distinct stress-free characteristics, in which the total strain field matches the spontaneous strain in the bulk domain, thereby minimizing the elastic strain. This process is physically equivalent to the adjustment of the simulation box size in an MD-NPT ensemble in response to lattice distortion, where the system releases internal stress by modifying its macroscopic dimensions, leading to a zero average stress level. In the vicinity of the domain wall center, the spontaneous strain components undergo drastic variations due to the rapid inversion of the polarization vector. However, governed by the displacement compatibility condition in continuum mechanics, the total strain must maintain spatial continuity and cannot perfectly track the local abrupt changes in $\epsilon^{sp}$. This local geometric mismatch forces the system to generate elastic compensation, manifested as a significant peak in the elastic strain at the DW center. This indicates that the PINN successfully captures the intrinsic mechanical mechanism of the DW as a "defect" structure that must sustain local elastic energy to maintain lattice integrity.

\begin{figure}[H]
    \centering
    \includegraphics[width=0.45\textwidth]{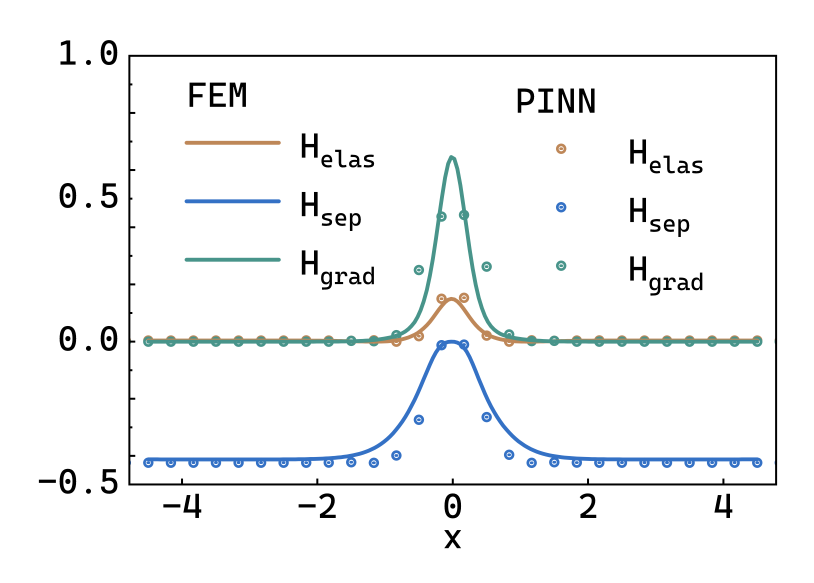}
    \caption{Energy densities output from PINN and FEM for the $180^\circ$ domain wall case.}
    \label{fig:180DWEnergyDensity}
\end{figure}

From an energetic perspective, as shown in Fig.~\ref{fig:180DWEnergyDensity}, under the stress-free condition, the elastic energy contribution remains relatively small and exhibits only a modest peak at the DW core, arising from the unavoidable local stress concentration. Therefore, in the stress-free state, most of the polarization-reversal cost is not related to elastic deformation, but instead primarily arises from the balance of separation and gradient energies.

\subsection{\texorpdfstring{Reconstruction of 90\textdegree{} Domain Wall Fields}{Reconstruction Fields of $90^\circ$ Domain Wall}}

\begin{figure}[H]
    \centering
    \includegraphics[width=0.45\textwidth]{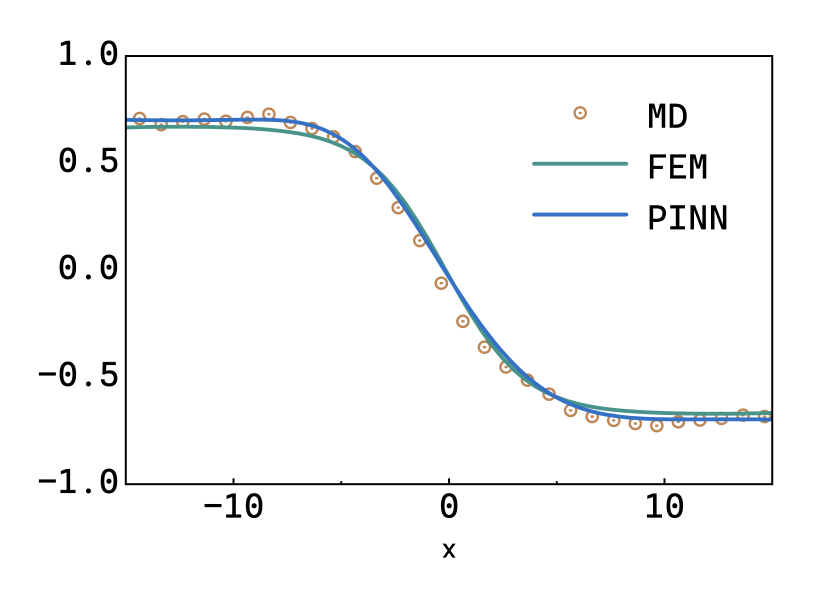}
    \caption{One-dimensional polarization profile of the $90^\circ$ domain wall}
    \label{fig:90DW_profile}
\end{figure}

Fig.~\ref{fig:90DW_profile} compares the one-dimensional polarization profiles obtained from MD, PINN, and FEM along the central line $y=z=0$ of the simulation box for a 90\textdegree{} DW. Similar to the $180^\circ$ DW case, the PINN prediction shows near-perfect agreement with the MD reference, consistent with the convergence behavior observed in the loss evolution. In contrast, the FEM profile appears slightly smoother, which can be attributed to the intrinsic regularization of the continuum formulation and numerical discretization, while still reproducing the key characteristics of the $90^\circ$ domain-wall transition.

\begin{figure}[H]
    \centering

    \begin{subfigure}[b]{0.45\textwidth}
        \centering
        \includegraphics[width=\linewidth]{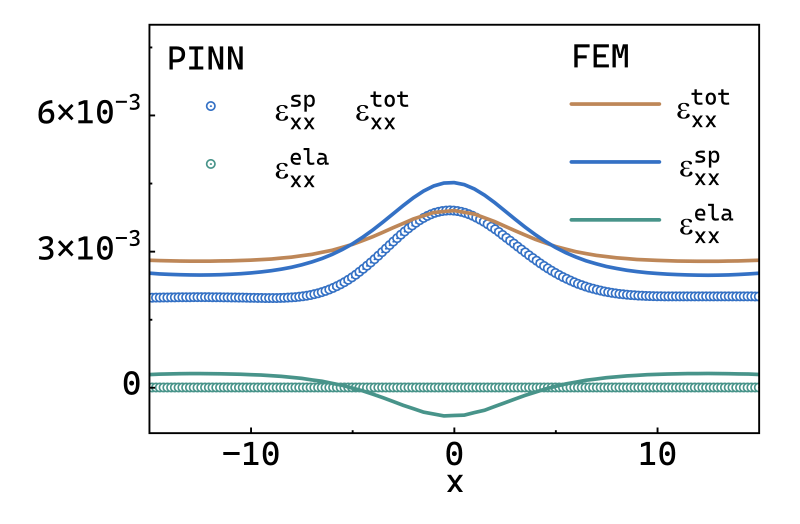}
        \caption{}
    \end{subfigure}
    \hspace{0.01\textwidth}
    \begin{subfigure}[b]{0.45\textwidth}
        \centering
        \includegraphics[width=\linewidth]{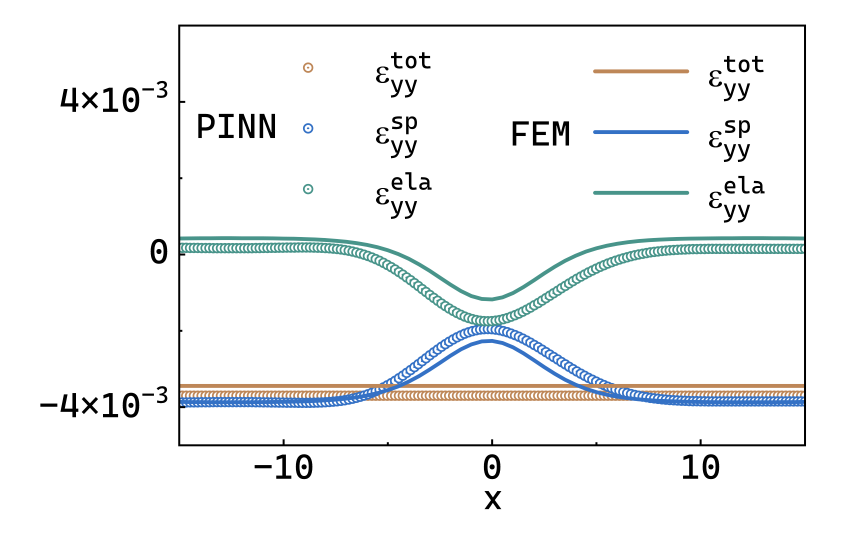}
        \caption{}
    \end{subfigure}
    \hspace{0.01\textwidth}
    \begin{subfigure}[b]{0.45\textwidth}
        \centering
        \includegraphics[width=\linewidth]{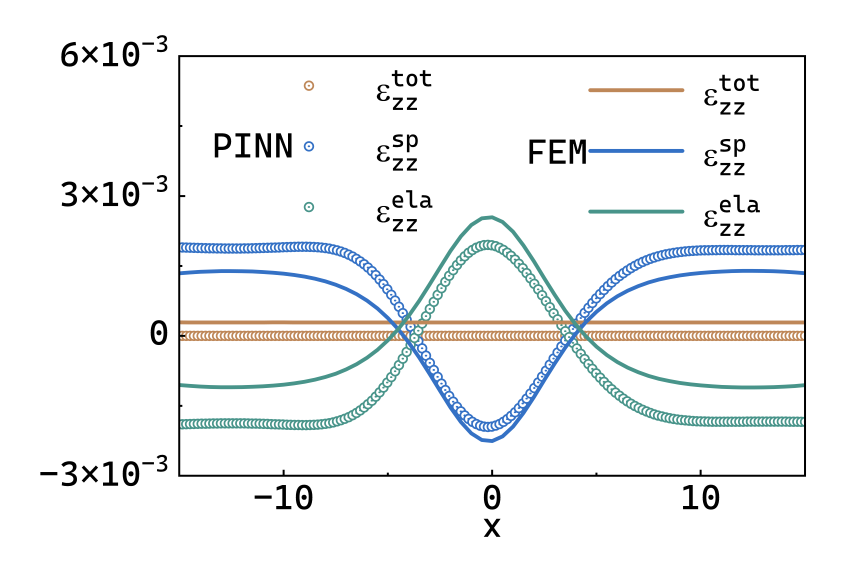}
        \caption{}
        \label{fig:c}
    \end{subfigure}
    \hspace{0.01\textwidth}
    \begin{subfigure}[b]{0.45\textwidth}
        \centering
        \includegraphics[width=\linewidth]{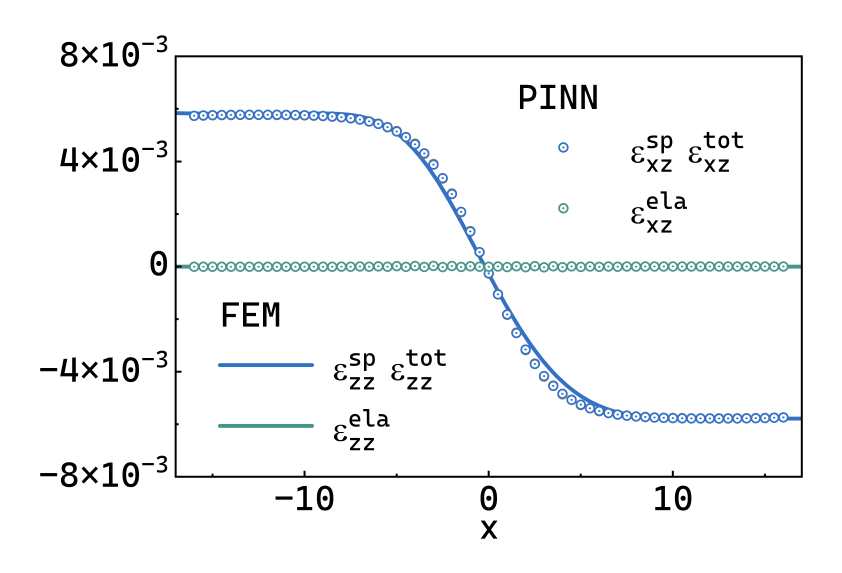}
        \caption{}
        \label{fig:d}
    \end{subfigure}
    \hspace{0.01\textwidth}
    \begin{subfigure}[b]{0.45\textwidth}
        \centering
        \includegraphics[width=\linewidth]{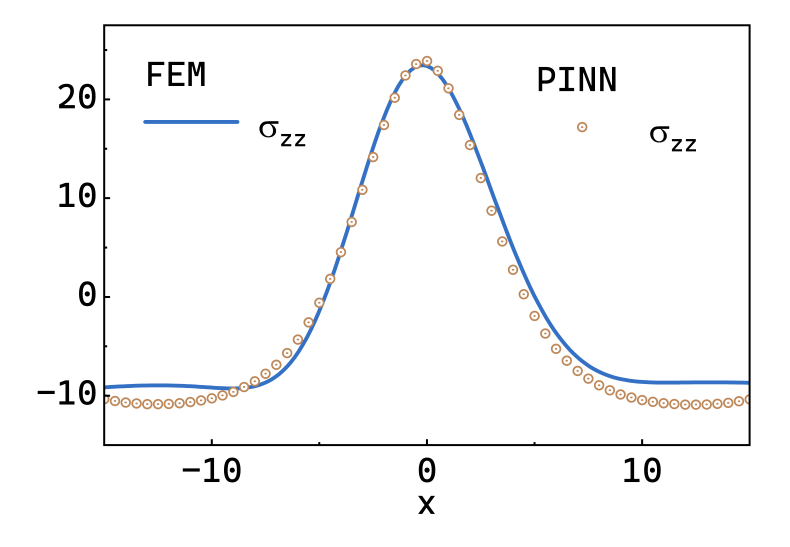}
        \caption{}
        \label{fig:e}
    \end{subfigure}

    \caption{Strain/stress fields output from both PINN and FEM of $90^\circ$ DW case: 
    (a) xx, 
    (b) yy, 
    (c) zz, 
    (d) xz,
    (e) stress along zz.}
    \label{fig:90DWPINNStrainStressField}
\end{figure}

As shown in Fig.~\ref{fig:90DWPINNStrainStressField}, the strain and stress fields predicted by the PINN show good agreement with the FEM reference for the $90^\circ$ DW case. Similar to the $180^\circ$ DW configuration, these results are consistent with the stress-free setting motivated by the MD-NPT ensemble, in which the system can relax its macroscopic deformation to release internal stress. In the saturated domain regions far from the wall, the total normal strains closely follow the corresponding spontaneous strains, so that the elastic parts remain small and the stress level stays near zero. 
In the vicinity of the $90^\circ$ DW center, the polarization rotates between two ferroelastic variants, leading to pronounced variations of the spontaneous strain components. In particular, the coupling-induced changes in $\epsilon_{xx}^{sp}$, $\epsilon_{yy}^{sp}$, and $\epsilon_{zz}^{sp}$ are reflected by clear spatial modulations around the wall, while the shear component $\epsilon_{xz}^{sp}$ exhibits a characteristic sign-changing transition across the DW, consistent with the underlying ferroelastic switching. However, due to the displacement compatibility constraint in continuum mechanics, the total strain field must remain continuous and cannot exactly reproduce the local abrupt variation of $\epsilon^{sp}$ at the DW core. As a consequence, an elastic compensation strain is generated to accommodate the geometric mismatch, which manifests as localized deviations between $\epsilon^{tot}$ and $\epsilon^{sp}$ and produces a non-negligible elastic strain concentration near the wall. This elastic accommodation results in a localized stress response, as evidenced by the peak of $\sigma_{zz}$ at the DW center in Fig.~\ref{fig:90DWPINNStrainStressField}(e).

\begin{figure}[H]
    \centering
    \includegraphics[width=0.5\textwidth]{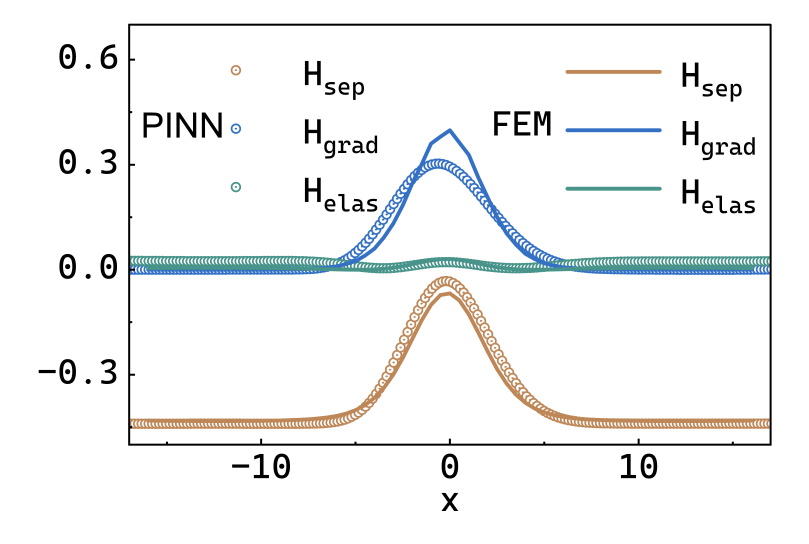}
    \caption{Energy density from PINN and FEM for the $90^\circ$ domain wall case.}
    \label{fig:90DWPINNEnergyDensity}
\end{figure}

Fig.~\ref{fig:90DWPINNEnergyDensity} compares the energy density distributions predicted by the PINN against the FEM reference for the $90^\circ$ DW case. For this configuration, the elastic energy density is nearly negligible across the entire domain and remains close to zero even in the vicinity of the wall, indicating that the stress-free relaxation effectively eliminates the elastic penalty associated with the ferroelastic switching. Consequently, the domain-wall energy is dominated by the competition between the separation energy density $H_{\mathrm{sep}}$ and the gradient energy density $H_{\mathrm{grad}}$. Specifically, $H_{\mathrm{grad}}$ exhibits a pronounced peak at the DW center, reflecting the strong polarization-gradient localization during the $90^\circ$ polarization rotation, whereas $H_{\mathrm{sep}}$ provides a negative contribution that partially offsets the gradient penalty and stabilizes the wall structure. Overall, the excellent agreement between PINN and FEM for all energy components confirms that the proposed framework accurately captures the energetic balance governing the $90^\circ$ DW, where the wall profile is primarily set by gradient regularization and separation driving, with only a minimal contribution from elastic accommodation.

\subsection{Domain Wall Properties}

\begin{table}[H]
\captionsetup{justification=centering, skip=4pt}
\caption{Comparsions of DW properties between MD and the current work}
\label{tab:compar_dw_properties}
\centering
\begin{tabular}{lll@{\hspace{2em}}lll}
\toprule
\textbf{Parameters} & \textbf{MD} & \textbf{PINN} & \textbf{FEM}
& \textbf{Unit} \\
\midrule
$l_{180}$ & 0.46 & 0.447 & 0.43 & nm \\
$l_{90}$  & 3.16  & 3.27  & 2.9 & nm \\
$\gamma_{180}$ & 4.8 & 4.8 & 4.64 & $\mathrm{mJ/m^2}$ \\
$\gamma_{90}$ & 4.0  & 12 & 10.78 & $\mathrm{mJ/m^2}$ \\
\bottomrule
\end{tabular}
\end{table}

The accuracy of the results obtained in this study is verified by comparing the DW width and the DW energy per unit area, as calculated by the proposed PINN and FEM, with those obtained from MD simulations. The comparison is summarized in Table~\ref{tab:compar_dw_properties}. For the fitting of the DW width, a hyperbolic tangent function is employed. We extracted a one-dimensional profile of $P_z$ along the direction of polarization switching direction in the global coordinate for both $180^\circ$ and $90^\circ$ DWs. The fitting formula is given as
\[
P(x) = P_{\text{effective}} \tanh\left( \frac{2(x - x_{\text{center}})}{l_{\text{width}}} \right)
\]
where $x_{\text{center}}$ denotes the position of the DW center, corresponding to the location where $P_z \approx 0$; $l_{\text{width}}$ represents the DW width; and $P_{\text{effective}}$ is the effective polarization amplitude, which is equal to $P_s$ for $180^\circ$ DW and $P_s/\sqrt{2}$ for $90^\circ$ DW. In addition, the DW energy is evaluated using the Eq.~[\ref{eq:dw_energy_1d}] in Section~\ref{sec:pinn}.

As summarized in Table~\ref{tab:compar_dw_properties}, the proposed PINN framework yields DW widths that are in good overall agreement with the MD reference for both the $180^\circ$ and $90^\circ$ DWs. The agreement is particularly encouraging for the broader $90^\circ$ DW, for which the PINN result remains closer to MD than the corresponding FEM prediction. It should also be noted that differences in the extracted DW width may partly originate from the time-averaging used to obtain the mean polarization profile in MD.

By contrast, the DW energy comparison requires more careful interpretation. The wall energies reported by PINN and FEM are obtained from continuum calculations based on the simulated field distributions, whereas the MD values are derived from the external work obtained by integrating the stress response during strain-driven transformation to a single-domain state. Because these quantities are not evaluated from an identical definition, Table~\ref{tab:compar_dw_properties} should be viewed primarily as a comparison of characteristic energetic scales rather than a strict one-to-one validation. Under this interpretation, the agreement for the $180^\circ$ DW remains satisfactory, while a more pronounced discrepancy is observed for the $90^\circ$ DW, for which both PINN and FEM predict higher energies than MD.

\section*{Conclusion}
In this paper, a Physics-Informed Neural Network (PINN) framework driven by molecular dynamics (MD) data is developed. The loss function of the network consists of two components: a supervised term that fits the discrete spatial polarization distributions obtained from MD simulations, and a physics-based term that incorporates the residuals of steady-state ferroelectric phase-field partial differential equations (PDEs). Since the overall objective involves multiple heterogeneous loss terms associated with different physical mechanisms, together with several strongly coupled trainable parameters, the training can be viewed as both a multi-task optimization problem and a coupled inverse-identification problem. To address the former, the GradNorm (GN) method is employed to adaptively balance the different loss components according to their gradient magnitudes, thereby preventing any single term from dominating the optimization process. To address the latter, a multi-stage training strategy is introduced, in which different parameter groups are activated and optimized sequentially to improve robustness and identifiability.

By minimizing the adaptively weighted total loss, the model not only reconstructs the polarization field along with the associated strain, stress, and energy density fields at the continuum scale, but also identifies critical physical parameters required for the phase-field model, including the characteristic energy density, characteristic length factor, gradient energy anisotropy factors, and Landau polynomial coefficients. By embedding the neural network-predicted physical parameters into COMSOL Multiphysics and solving the corresponding phase-field PDEs, we demonstrate that the learned parameters enable accurate reproduction of not only the ferroelectric domain structures but also the associated material responses, including built-in electric fields, stress/strain distributions, and the energy landscape. This framework provides an effective methodology for establishing multiscale connections between atomistic and continuum descriptions and holds the potential to infer underlying physical properties directly from polarization distributions of a wide range of materials.

\section*{Acknowledgments}
Xuejian Wang, Frank Wendler, Michael Zaiser and Xingchen Tan are grateful for financial support from the Deutsche Forschungsgemeinschaft (DFG) under grant GRK2495/2 K. This work was supported by Japan Society for the Promotion of Science Japanese-German Graduate Externship. H.A. is financially supported by JST SPRING (grant no. JPMJSP2112). Simulations were performed, in part, on Fujitsu PRIMEHPC FX1000 computer in Information Technology Center of Nagoya University through HPCI projects (grant nos.hp240100 and hp250132).

\appendix
\renewcommand{\thesection}{Appendix \Alph{section}}

\setcounter{figure}{0}
\setcounter{table}{0}
\renewcommand{\thefigure}{\thesection\arabic{figure}}
\renewcommand{\thetable}{\thesection\arabic{table}}

\section{Calculations of Elastic, Dielectric and Piezoelectric Constants in MD}
\label{app:md_constants}

\setcounter{figure}{0}
\setcounter{table}{0}

We prepared a supercell containing 16$^3$ unit cells. First, we performed N$\sigma$T-MD simulations at a temperature of 300 K and zero external stress, from which we obtained the lattice constants $a$ and $c$ as the average unit-cell dimensions, as well as the spontaneous polarization $P_s$~\cite{azuma2025unique}. The piezoelectric stress coefficients $e_{ij}$ were then evaluated from NVT-MD simulations performed at the fixed cell size corresponding to these lattice constants $a$ and $c$, by applying an external electric field of 5 MV/m along six directions ($+x$, $-x$, $+y$, $-y$, $+z$, $-z$) and computing the change in the time-averaged stress as a function of the applied field. The elastic constants were taken from Ref. \cite{azuma2025unique}, where they were obtained from the stress response to imposed strain in NVT-MD simulations. The dielectric permittivity was calculated from the variance of the total electric dipole moment $M$ of the system as
$$
k_a=\frac{\left<M_a^2\right>-\left<M_a\right>^2}{V k_B T}
$$
where $V$ is the supercell volume, $k_B$ is the Boltzmann constant, $\left<...\right>$ denotes a time average over the MD trajectory, $T$ is the simulation finite temperature range from 280 to 350$K$. The validity of this expression for predicting reasonable dielectric permittivities in BaTiO$_3$ has been demonstrated in \cite{Hashimoto2015MDpermittivity}.

\begin{figure}[H]
    \centering
    \includegraphics[width=0.45\textwidth]{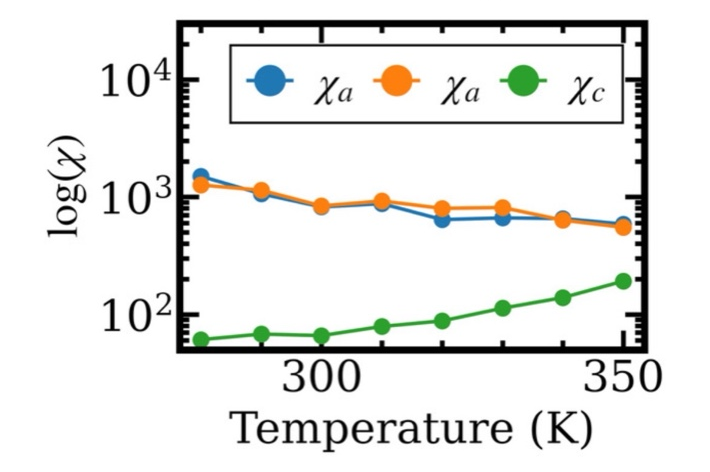}
    \caption{Temperature dependence of the dielectric susceptibilities obtained from MD simulations, resolved along directions parallel and perpendicular to the spontaneous polarization}
    \label{fig:md_xi}
\end{figure}

Fig.~\ref{fig:md_xi} shows the temperature dependence of the dielectric susceptibilities obtained from the MD simulations, resolved along directions parallel and perpendicular to the spontaneous polarization. A clear anisotropy is observed between the two components. The susceptibility along the polarization direction, $\chi_c$, increases markedly as the temperature approaches the transition region, exhibiting a characteristic Curie--Weiss--type~\cite{Cochran1960Ferroelectricity} enhancement associated with the soft ferroelectric mode. In contrast, the transverse component $\chi_a$ shows only a weak temperature dependence over the same temperature range, and even displays a slight decreasing trend. This behavior indicates that $\chi_a$ is dominated by the background dielectric response rather than by a soft-mode contribution. Based on these observations, different fitting strategies are adopted for the two components. The longitudinal susceptibility $\chi_c(T)$ is fitted using a Curie--Weiss--type expression,
\begin{equation}
\chi_c(T)=\chi_{c,\mathrm{bg}}+\frac{C_1}{T_c-T}
\end{equation}
in order to separate the background dielectric constant $\chi_{c,\mathrm{bg}}$ from the soft-mode contribution. For the transverse component, where no clear Curie--Weiss--type behavior is observed, the susceptibility is approximated as a weakly temperature-dependent background term,
\begin{equation}
\chi_a(T)\approx \chi_{a,bg}+C_2(T-T_0)
\end{equation}
or, in the simplest approximation, as a constant background value $\chi_{a,\mathrm{bg}}$.

The corresponding background dielectric constants are then obtained from the fitted susceptibilities through
\begin{equation}
k_i = \varepsilon_0 \left(1+\chi_{i,\mathrm{bg}}\right), \qquad i=a,c
\end{equation}
where \(k_i\) is the dielectric constant along the crystallographic direction \(i\), \(\chi_{i,\mathrm{bg}}\) is the fitted background susceptibility, and \(\varepsilon_0\) is the vacuum permittivity. In this way, the fitted values of \(\chi_{c,\mathrm{bg}}\) and \(\chi_{a,\mathrm{bg}}\) are converted into the dielectric constants ($k_a$ and $k_c$) listed in Table~\ref{tab:parameters}.

\begin{table}[H]
\centering
\captionsetup{skip=4pt}
\begin{threeparttable}
\caption{Material and model parameters for BTO at 300 K determined from MD data and used in the PINN and FEM frameworks}
\label{tab:parameters} 
\begin{tabular}{lll@{\hspace{2em}}lll}
\toprule
\textbf{Parameter} & \textbf{Value} & \textbf{Unit} 
& \textbf{Parameter} & \textbf{Value} & \textbf{Unit} \\
\midrule
$e_{31}$ & 0.33 & C\,m$^{-2}$ & $a$ & 0.3996 & nm \\
$e_{33}$ & 4.47 & C\,m$^{-2}$ & $c$ & 0.4043 & nm \\
$e_{15}$ & 15.7 & C\,m$^{-2}$ & $P_s$ & 0.18 & C\,m$^{-2}$ \\
$k_a$ & $7.7\times 10^{-9}$ & C(V\,m)$^{-1}$ 
& $\chi_{a,\mathrm{bg}}$ & 871 & (–) \\
$k_c$ & $1.0\times 10^{-9}$ & C(V\,m)$^{-1}$ 
& $\chi_{c,\mathrm{bg}}$ & 112 & (–) \\
$C_{11}^{T}$ & 200 & GPa & $C_{12}^{T}$ & 110 & GPa \\
$C_{13}^{T}$ & 100 & GPa & $C_{33}^{T}$ & 120 & GPa \\
$C_{44}^{T}$ & 50 & GPa & $C_{66}^{T}$ & 140 & GPa \\
$l_{180}^{MD}$ & 0.446 & nm & $l_{90}^{MD}$ & 3.16 & nm \\
$\gamma_{180}^{MD}$ & 4.8 & mJ\,m$^{-2}$ & $\gamma_{90}^{MD}$ & 4.2 & mJ\,m$^{-2}$ \\
$\beta_0$ & 0.9 & kA\,(V\,m)$^{-1}$ & $\eta$ & 4/3 & (–) \\
\bottomrule
\end{tabular}
\begin{tablenotes}
\footnotesize 
\item Note: The superscript $T$ denotes tetragonal elastic tensor components obtained from MD.
\end{tablenotes}
\end{threeparttable}
\end{table}

\section{Preprocessing of MD Data}
\label{app:md_preprocessing}

\setcounter{figure}{0}
\setcounter{table}{0}
\renewcommand{\thefigure}{\thesection\arabic{figure}}
\renewcommand{\thetable}{\thesection\arabic{table}}

Before executing PINN training and FEM calculations, the MD data need to be preprocessed so that they can be directly used as inputs for both the PINN and FEM frameworks. The data provided by MD can be divided into two categories. One consists of the material parameters directly obtained from MD, such as those listed in Table~\ref{tab:parameters}, The other consists of the discrete distribution of polarization, which needs to be transformed into a structured dataset suitable for PINN training. 

\paragraph{Non-dimensionialization}

To ensure dimensional consistency among the governing equations, constitutive relations, and loss terms used in the PINN training, all variables and material parameters are nondimensionalized by introducing the following characteristic scales:
\begin{equation}
P_{\mathrm{scale}} = P_s,
\qquad
e_{\mathrm{scale}} = P_s,
\qquad
D_{\mathrm{scale}} = P_{\mathrm{scale}},
\qquad
\gamma_{\mathrm{scale}} = \gamma_{180}^{\mathrm{MD}},
\qquad
l_{\mathrm{scale}} = 0.4~\mathrm{nm},
\label{eq:scale_basic}
\end{equation}
and
\begin{equation}
G_{\mathrm{scale}} = \frac{\gamma_{\mathrm{scale}}}{l_{\mathrm{scale}}},
\qquad
C_{\mathrm{scale}} = G_{\mathrm{scale}},
\qquad
E_{\mathrm{scale}} = \frac{G_{\mathrm{scale}}}{P_{\mathrm{scale}}},
\qquad
K_{\mathrm{scale}} = \frac{D_{\mathrm{scale}}}{E_{\mathrm{scale}}}.
\label{eq:scale_derived}
\end{equation}

Accordingly, the resulting nondimensional material parameters used in the present work are summarized in Table~\ref{tab:parameters_nd}.

\begin{table}[H]
\centering
\captionsetup{skip=4pt}
\begin{threeparttable}
\caption{Nondimensionalized material and model parameters for BTO at 300~K.}
\label{tab:parameters_nd}

\begin{tabular}{llllll}
\toprule
\textbf{Parameter} & \textbf{Value} & \textbf{Parameter} & \textbf{Value} & \textbf{Parameter} & \textbf{Value} \\
\midrule
$e_{31}$   & 1.83   & $C_{11}$  & 14416 & $l_{180}^{MD}$         & 1.125 \\
$e_{33}$   & 24.83  & $C_{12}$  & 8583  & $l_{90}^{MD}$          & 7.9 \\
$e_{15}$   & 87.2   & $C_{44}$  & 3750  & $\gamma_{180}^{MD}$    & 1 \\
$k_a$      & 2.86   & $k_c$     & 0.37  & $\gamma_{90}^{MD}$     & 0.83 \\
$P_s$      & 1      &           &       &                        &   \\
\bottomrule
\end{tabular}

\begin{tablenotes}
\footnotesize
\item Note: The elastic constants $C_{11}$, $C_{12}$, and $C_{44}$ listed here are effective cubic elastic constants obtained in Section~\ref{sec:pfm}.
\end{tablenotes}

\end{threeparttable}
\end{table}

\paragraph{Preprocessing of polarization profiles}
The MD data processing workflow aims to convert discrete MD data into a continuous representation that can be recognized by neural networks based on PFM. The whole workflow can be summarized as follows:

\begin{figure}[H]
    \centering
    \includegraphics[width=0.35\textwidth]{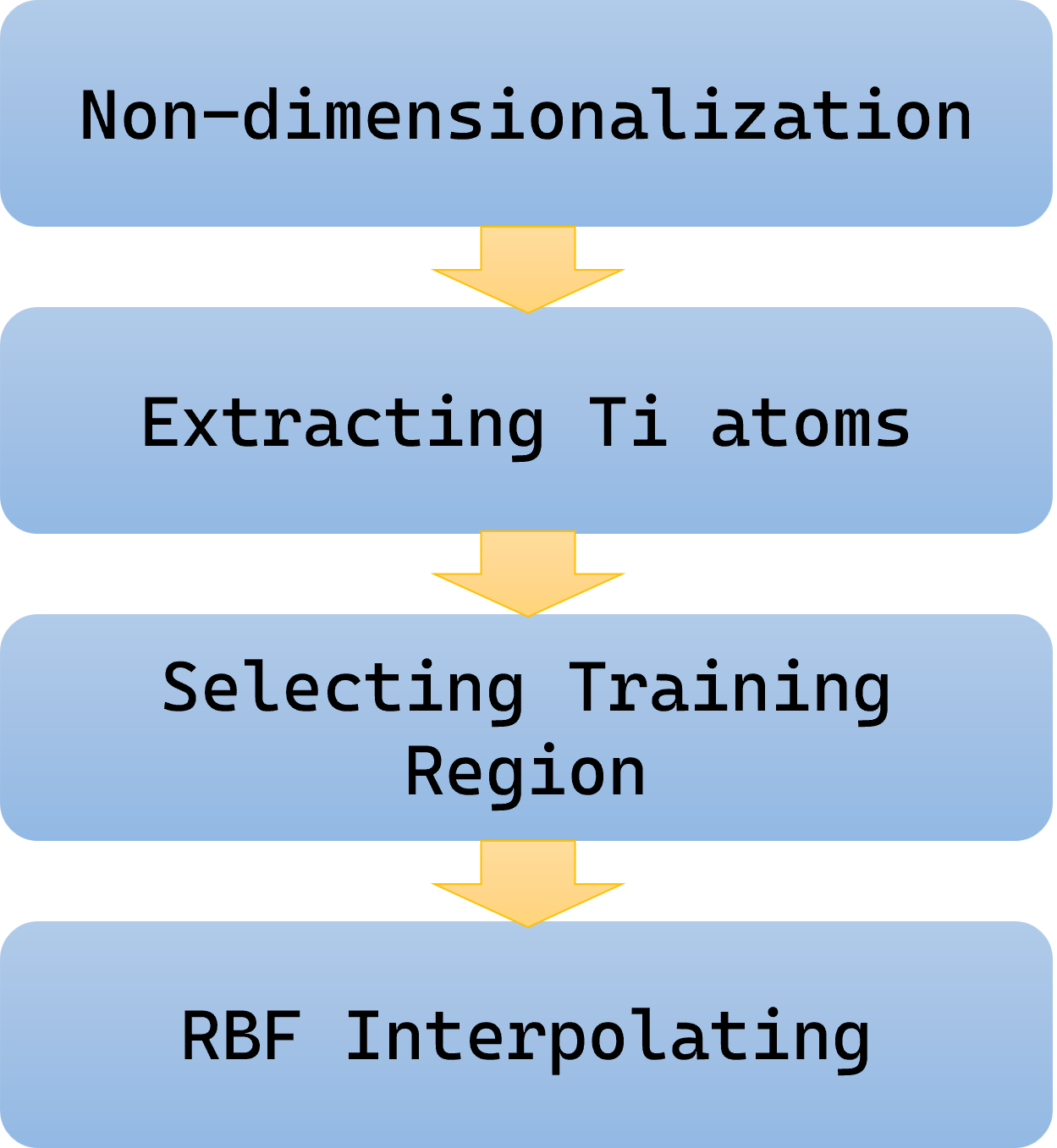}
    \caption{Workflow of GradNorm method}
    \label{fig:MDdataprocress}
\end{figure}

The MD datasets of both 180$^\circ$ and 90$^\circ$ DWs are preprocessed through a unified pipeline, as illustrated in Fig.~\ref{fig:MDdataprocress}. First, all atomic coordinates and polarization vectors are non-dimensionalized using characteristic scales. Only Titanium (Ti) atoms are retained, as spontaneous polarization in BaTiO$_3$ predominantly originates from Ti displacements within the oxygen octahedral cages. A central subregion of the simulation box is selected to mitigate boundary artifacts and focus on the intrinsic DW structure.

Polarization vectors $\mathbf{P}_i$ are first computed for each Ti-centered local unit cell using the microscopic dipole formula,
\begin{equation}
\mathbf{P}_i
= \frac{1}{V_{\mathrm{cell}}}
\sum_{a} q_a \bigl(\mathbf{r}_a - \mathbf{r}_{\mathrm{Ti}}\bigr),
\label{eq:local_P}
\end{equation}
where the summation runs over one Ti atom, its six nearest O atoms, and eight Ba atoms. The atomic positions $\mathbf{r}_a$ are taken from the MD configuration relative to the ideal cubic reference structure.

The resulting discrete polarization vectors at Ti positions $\mathbf{r}_i$ are then interpolated onto a regular three-dimensional grid using a radial basis function (RBF) approximation,
\begin{equation}
\mathbf{P}(\mathbf{r}) =
\sum_{i=1}^{N} 
\mathbf{P}_i \,
\phi\!\left(\left|\mathbf{r}-\mathbf{r}_i\right|\right),
\label{eq:RBF_interp}
\end{equation}
where $\phi(\cdot)$ is a chosen scalar RBF kernel (e.g., multiquadric or thin-plate spline). This procedure yields a smooth continuum polarization field $\mathbf{P}(\mathbf{r})$ consistent with the atomistic MD data.
This preprocessing pipeline yields smooth, structured polarization fields, suitable for subsequent modeling and analysis of both 180$^\circ$ and 90$^\circ$ DWs behaviors.

\begin{figure}[H]
    \centering

    \begin{subfigure}[b]{0.45\textwidth}
        \centering
        \includegraphics[width=\linewidth]{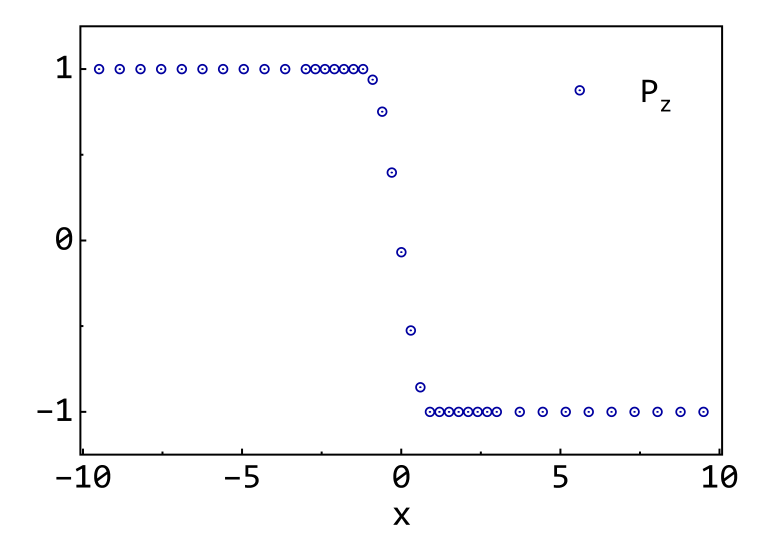}
        \caption{}
    \end{subfigure}
    \hspace{0.01\textwidth}
    \begin{subfigure}[b]{0.45\textwidth}
        \centering
        \includegraphics[width=\linewidth]{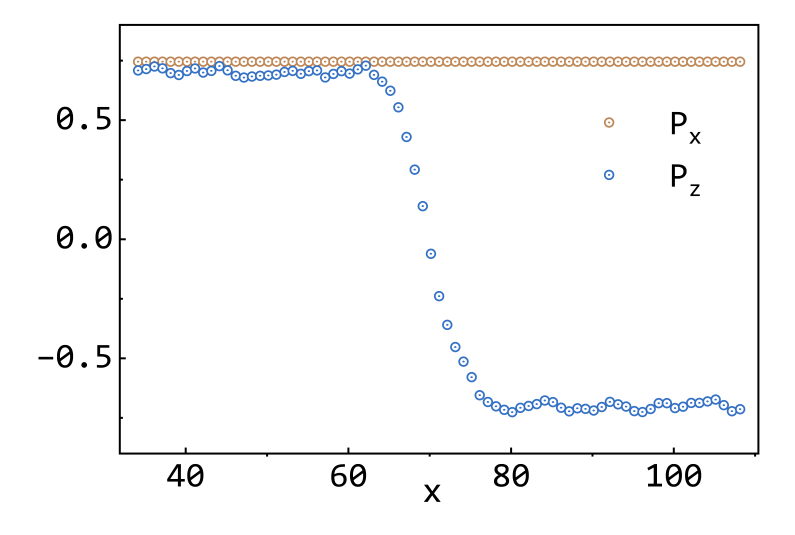}
        \caption{}
    \end{subfigure}
  
    \caption{1D Distribution of polarization calculated by MD and preprocressed by the workflow:
    (a) $180^\circ$ DW, 
    (b) $90^\circ$ DW.}
    \label{fig:MD_Polarization}
\end{figure}

Fig.~\ref{fig:MD_Polarization} displays the 1D polarization profile of $180^\circ$ and $90^\circ$ calculated from MD with the above data preprocessing. It can be observed that the MD simulations accurately reproduce the characteristic features of both the $180^\circ$ and $90^\circ$ DWs. However, compared to the $180^\circ$ DW case, the $90^\circ$ DW exhibits more pronounced fluctuations.

\begin{figure}[H]
    \centering
    \includegraphics[width=0.45\textwidth]{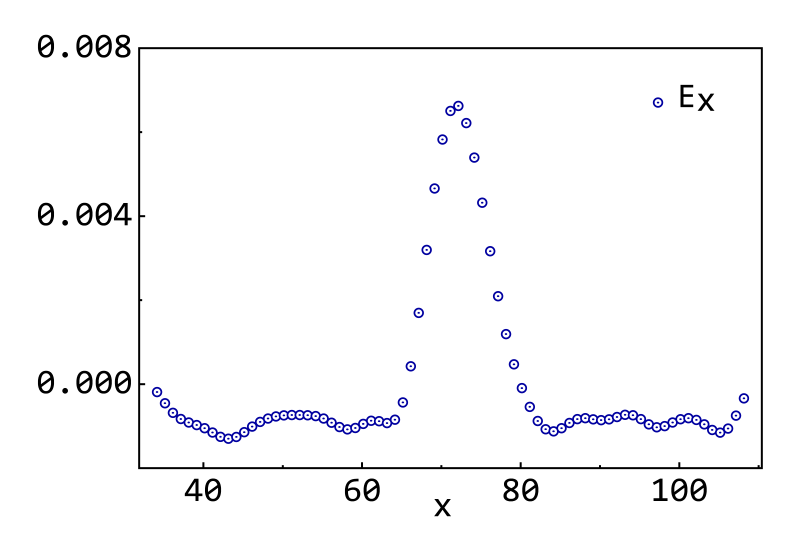}
    \caption{1D Distribution of electric field of the $90^\circ$ DW calculated by MD and preprocressed by the workflow}
    \label{fig:90ElecField}
\end{figure}

In addition to the polarization field, the electrostatic response from MD is also incorporated into the PINN as a known quantity.
Specifically, the discrete MD electrostatic potential values $\{\phi_i\}$ sampled at Ti positions $\{\mathbf{r}_i\}$ are first non-dimensionalized and then interpolated onto the same regular grid as $\mathbf{P}(\mathbf{r})$ using the RBF approximation,
\begin{equation}
\phi(\mathbf{r}) =
\sum_{i=1}^{N} 
\phi_i \,
\phi_{\mathrm{RBF}}\!\left(\left|\mathbf{r}-\mathbf{r}_i\right|\right),
\label{eq:RBF_interp_phi}
\end{equation}
where $\phi_{\mathrm{RBF}}(\cdot)$ denotes the selected scalar RBF kernel and the interpolation settings are consistent with those used for the polarization field.

To mitigate the amplification of high-frequency numerical noise during differentiation, the interpolated potential is further smoothed by a Gaussian filter,
\begin{equation}
\phi_s(\mathbf{r}) = \mathcal{S}_{\sigma}\!\left[\phi(\mathbf{r})\right],
\label{eq:phi_smooth_md}
\end{equation}
where $\mathcal{S}_{\sigma}$ is a Gaussian smoothing operator with $\sigma$ defined in grid units.
The electric field is then computed by numerical differentiation on the regular grid,
\begin{equation}
\mathbf{E}(\mathbf{r}) = -\nabla \phi_s(\mathbf{r}),
\label{eq:E_from_phi_md}
\end{equation}
with the grid spacing used in the finite-difference evaluation of $\nabla$.

The resulting field $\mathbf{E}(\mathbf{r})$ is finally extracted along the center line of the domain-wall plane to obtain a 1D profile, as shown in Fig.~\ref{fig:90ElecField}.

\section{Workflow of GradNorm Method}
\label{app:gradnorm}

\setcounter{figure}{0}
\setcounter{table}{0}
\renewcommand{\thefigure}{\thesection\arabic{figure}}
\renewcommand{\thetable}{\thesection\arabic{table}}

In this study, the GradNorm method~\cite{chen2018gradnorm} is used to drive all tasks to converge at a similar training speed. This is achieved by continuously adjusting the task weights $w_i$ such that the gradient norm $G\!N_i$ of each task approaches a target value 
$g_{\mathrm{target},i}$. Here, $L_{\mathrm{GL}}$, $L_{\mathrm{BL,elec}}$, and $L_{\mathrm{BL,mech}}$ are the loss terms involved in the GradNorm (GN) procedure, and $w_i$ denotes the adaptive weight associated with each term. This basic idea is implemented through the following steps:

\paragraph{Temperature-based weight normalization}
In the original GradNorm formulation
~\cite{chen2018gradnorm}, the task weights are directly used in the weighted loss without any additional normalization. However, in our tests, we found that the task weights easily become negative and change too rapidly during training. Therefore, to improve numerical stability, the raw task weights were further transformed into effective weights through a temperature-scaled softmax function, following a similar normalization approach as used in predictive uncertainty calibration~\cite{Guo2017Calibration}
\begin{equation}
\tilde{w}_i =
\frac{\exp\left(w_i^{\mathrm{raw}}/T\right)}
{\sum_{j=1}^{N}\exp\left(w_j^{\mathrm{raw}}/T\right)},
\end{equation}
where $T$ is a temperature-like coefficient controlling the smoothness of the normalized weight distribution. After several trials, $T$ was set to $0.15$ in the present work.

To ensure that the average task weight equals one, a normalization step is applied:
\begin{equation}
w_i = N \cdot \tilde{w}_i,
\end{equation}

where $N$ is the number of the GN training tasks. Thus, the normalized weights satisfy
\begin{equation}
\frac{1}{N}\sum_{i=1}^{N} w_i = 1.
\end{equation}

\paragraph{Gradient magnitude evaluation}
Secondly, we compute the gradient magnitude of each task. The base gradient norm is defined as
\begin{equation}
G_{i,\mathrm{base}} = 
\left\| 
\frac{\partial L_i}{\partial \boldsymbol{\Theta}} 
\right\|,
\end{equation}
which represents the natural influence of the $i$-th task on the shared parameters. $\Theta$ here is the neural network parameters(weight and bias), as well as the training PFM parameters ($G$, $l$, $\mu$ and Landau coefficients). 

After applying the task weights,
\begin{equation}
G\!N_i = w_i \cdot G_{i,\mathrm{base}}
\end{equation}
represents the gradient norm 
produced by the $i$-th task on the network.

\paragraph{Training-rate estimation and target gradient computation}
After starting the GradNorm algorithm, the initial loss of each task is recorded as $L_i(0)$. At training step $t$, 
the relative loss is defined as
\begin{equation}
r_i' = \frac{L_i(t)}{L_i(0)}
\end{equation}
and the relative training speed factor is evaluated as
\begin{equation}
r_i =
\frac{(r_i')^{\alpha}}
{\frac{1}{N}\sum_{j=1}^{N}(r_j')^{\alpha}},
\end{equation}
where $\alpha$ is a GradNorm hyperparameter.
Next, the training speed factors are multiplied with the average gradient magnitude
\begin{equation}
\bar{G\!N} = \frac{1}{N}\sum_{i=1}^{N} G\!N_i,
\end{equation}
to obtain target gradient magnitudes 
\begin{equation}
g_{\mathrm{target},i} = \bar{G\!N} \cdot r_i.
\end{equation}

\paragraph{GradNorm weight update and network optimization}
GradNorm aims to match the gradient norm
to the target value
\begin{equation}
G\!N_i \approx g_{\mathrm{target},i}.
\end{equation}
The GradNorm loss is
\begin{equation}
L_{\mathrm{GN}} =
\sum_{i=1}^{N}
\left(
G\!N_i - 
\mathrm{stopgrad}(g_{\mathrm{target},i})
\right)^2
\end{equation}
Here, $\mathrm{stopgrad}(\cdot)$ denotes the stop-gradient operation, meaning that its argument is treated as a constant during backpropagation. In this way, the target gradient norm $g_{\mathrm{target},i}$ serves only as a fixed reference for updating the adaptive weights, rather than being differentiated through the GradNorm loss itself.

The raw task weights are then updated
\begin{equation}
w_i^{\mathrm{raw}} \leftarrow
w_i^{\mathrm{raw}} -
\eta_w
\frac{\partial L_{\mathrm{GN}}}
{\partial w_i^{\mathrm{raw}}},
\end{equation}
and finally the neural network parameters and training parameters used in PFM are updated using the total weighted loss

\begin{equation}
\mathbf{\Theta} \leftarrow
\mathbf{\Theta} -
\eta_{\Theta}
\frac{\partial L_{\mathrm{total}}}
{\partial \mathbf{\Theta}}
\end{equation}

where $\eta_{\Theta}$ is learning rate of each parameter.

\section{Multi-stage Training Strategy}
\label{app:3step}

\setcounter{figure}{0}
\setcounter{table}{0}
\renewcommand{\thefigure}{\thesection\arabic{figure}}
\renewcommand{\thetable}{\thesection\arabic{table}}

\begin{figure}[H]
    \centering
    \includegraphics[width=0.6\textwidth]{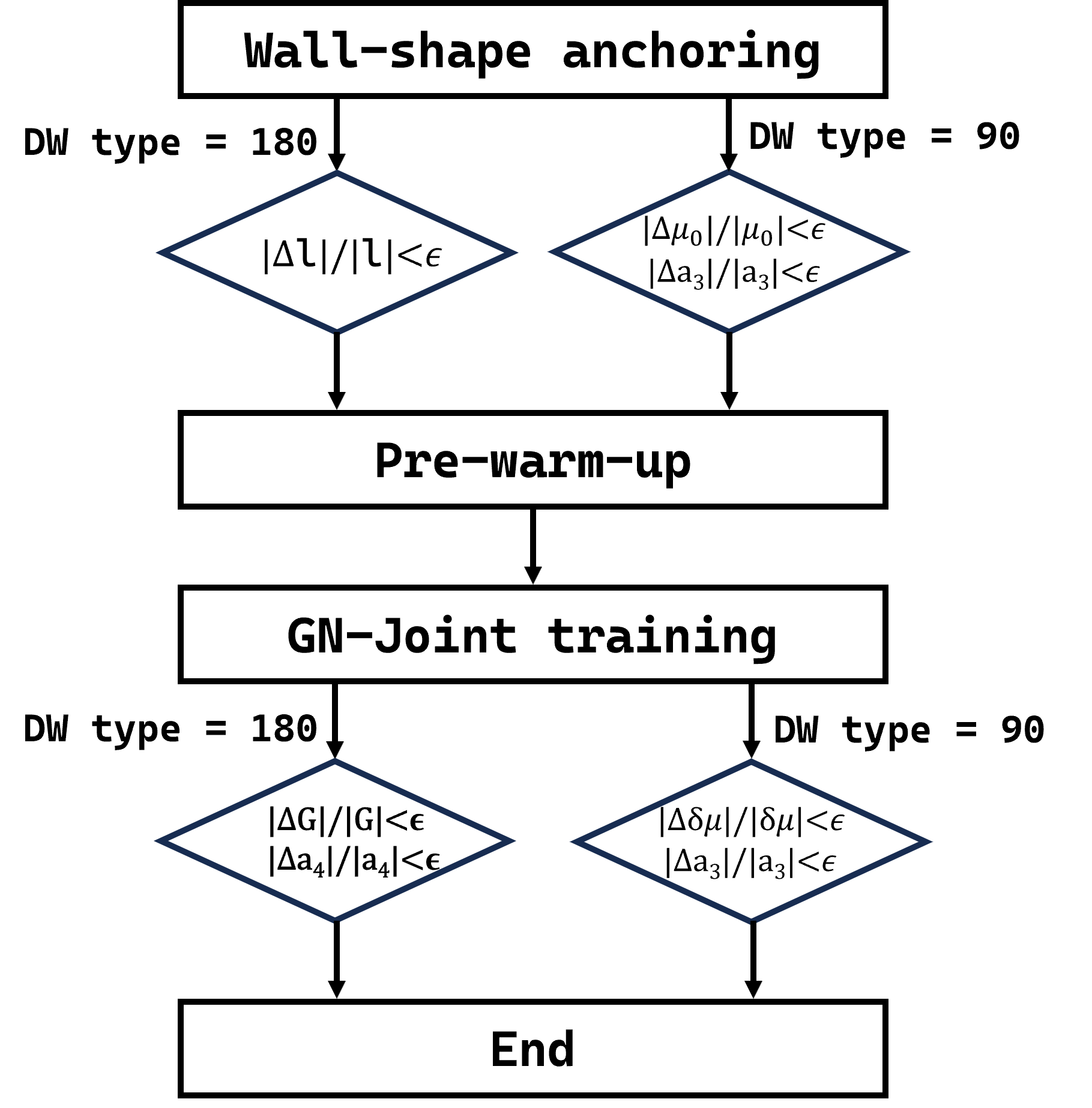}
    \caption{Multi-stage workflow for both $180^\circ$ and $90^\circ$ DW training}
    \label{fig:stageflow}
\end{figure}

To address the optimization issues discussed in Section~\ref{sec:pinn}, a multi-stage training strategy is introduced for both $180^\circ$ and $90^\circ$ DW cases. As shown in Fig.~\ref{fig:stageflow}, the strategy is divided into three stages. Overall, we define the relative parameter-update indicators as
\begin{equation}
\delta_\Theta=\frac{|\Delta \Theta|}{|\Theta|+\epsilon},
\label{eq:delta_Theta}
\end{equation}

where $\Theta$ are training parameters, and $\Delta \Theta$ denotes the parameter changes between two successive monitoring steps, and $\epsilon$ is a small positive constant introduced to avoid numerical singularity when the denominator becomes smaller than $\epsilon$. In this study, $\epsilon$ is set as $1\times10^{-4}$.

\paragraph{Stage 1 (wall-shape anchoring with simplified GL loss)}

In the first stage, the optimization aims to stabilize the DW profile and determines part of parameters. For the $180^\circ$ case, assuming $P_z$ switching along $x$ direction, $P_x$ and $P_y$ remain zero, and the full one-dimensional Ginzburg--Landau (GL) residual can be expressed as
\begin{equation}
\mathcal{R}^{GL,180}
=
G\frac{l^2}{P_s^2}\,\frac{\partial^2 P_z(x)}{\partial x^2}
-
G\frac{\partial \psi(0, 0, P_z(x))}{\partial P_z(x)}
-
\frac{\partial H_{bulk}(0, 0, P_z)}{\partial P_z(x)}
\label{eq:GL_residual_180_1d}
\end{equation},

which can be simplified by ignoring bulk terms and canceling $G$ to obtain 

\begin{equation}
\mathcal{R}^{180}_{\mathrm{GL_1}}
=
\frac{l^2}{P_s^2}\,\frac{\partial^2 P_z}{\partial x^2}
-
\frac{\partial \psi(0, 0, P_z)}{\partial P_z}.
\label{eq:GL_simplified_residual_180_1d}
\end{equation},

which is used in the first stage for $180^\circ$ case for anchoring the DW shape and $l$.

For the $90^\circ$ case, we also assume that $P_z$ switches along the $x$ direction in the global frame, while $P_x$ remains constant and $P_y$ is set to zero. Therefore, the GL equation can be simplified in the global frame by neglecting the bulk terms and canceling $G$
\begin{equation}
\begin{split}
\frac{l^2}{2P_s^2}(\mu+2)\frac{d^2P_z}{dx^2}
={}&
\frac{2a_1}{P_s^2}P_z
+
\frac{a_2}{2P_s^4}\left(4P_z^3+12P_x^2P_z\right)
\\
&+
\frac{a_3}{4P_s^4}\left(4P_z^3-4P_x^2P_z\right)
+
\frac{a_4}{4P_s^6}\left(30P_x^4P_z+60P_x^2P_z^3+6P_z^5\right).
\end{split}
\label{eq:GL_90_reduced}
\end{equation}

To reduce the strong coupling between $\mu$ and $a_3$, we introduce an auxiliary parameter $\mu_0$ and define
\begin{equation}
\frac{l^2}{2P_0^2}(\mu_0+2)\frac{d^2P_z}{dx^2}
=
\frac{2a_1}{P_0^2}P_z
+
\frac{a_2}{2P_0^4}\left(4P_z^3+12P_x^2P_z\right)
+
\frac{a_4}{4P_0^6}\left(30P_x^4P_z+60P_x^2P_z^3+6P_z^5\right),
\label{eq:GL_90_mu0}
\end{equation}
which is used to anchor $\mu_0$ in the first stage.

Subtracting Eq.~\eqref{eq:GL_90_mu0} from Eq.~\eqref{eq:GL_90_reduced}, and using $\mu=\mu_0+\Delta\mu$, yields
\begin{equation}
\frac{a_3}{4P_0^4}\left(4P_z^3-4P_x^2P_z\right)
=
\frac{l^2}{2P_0^2}\Delta\mu\frac{d^2P_z}{dx^2},
\label{eq:GL_90_delta_mu_a3}
\end{equation}
which is then used in the second stage to constrain $a_3$ while refining the correction term $\Delta\mu$.

Based on the derivation, Stage 1 is further divided into two substages, denoted as Stage 1(a) and Stage 1(b). In Stage 1(a), the residual associated with Eq.~\eqref{eq:GL_90_mu0} is defined as
\begin{equation}
\mathcal{R}^{90}_{\mathrm{GL_1(a)}}
=
\frac{l^2}{2P_0^2}(\mu_0+2)\frac{d^2P_z}{dx^2}
-
\frac{2a_1}{P_0^2}P_z
-
\frac{a_2}{2P_0^4}\left(4P_z^3+12P_x^2P_z\right)
-
\frac{a_4}{4P_0^6}\left(30P_x^4P_z+60P_x^2P_z^3+6P_z^5\right),
\label{eq:R_GL1a_90}
\end{equation}
which is used to anchor $\mu_0$ in the first substage.

In Stage 1(b), the residual associated with Eq.~\eqref{eq:GL_90_delta_mu_a3} is defined as
\begin{equation}
\mathcal{R}^{90}_{\mathrm{GL_1(b)}}
=
\frac{l^2}{2P_0^2}\Delta\mu\frac{d^2P_z}{dx^2}
-
\frac{a_3}{4P_0^4}\left(4P_z^3-4P_x^2P_z\right),
\label{eq:R_GL1b_90}
\end{equation}
where
\begin{equation}
\Delta\mu=\mu-\mu_0.
\label{eq:mu_split_90}
\end{equation}
This residual is used in the second substage to constrain $a_3$ while refining the correction term $\Delta\mu$.

The corresponding loss of Stage 1  could be expressed for both $180^\circ$ and $90^\circ$ DWs
\begin{equation}
L_{\mathrm{Step1}}
=
w_{\mathrm{data}} L_{\mathrm{data}}
+
w_{\mathrm{GL}} \left\| \mathcal{R}^{i}_{\mathrm{GL}} \right\|^2,
\label{eq:loss_step1}
\end{equation}

where $i$ stands for 180 or 90.

\paragraph{Stage 2 (warmup-based coupled-parameter release)}
In the second stage, a linear warm-up schedule is introduced to prevent abrupt gradient oscillations and to ensure a smoother transition when the newly added loss term becomes active.

A linear warmup factor $\alpha(it)$ is introduced to gradually activate the corresponding loss contribution
\begin{equation}
\alpha(it)=
\begin{cases}
0, & it < it_{\mathrm{stage1,end}}, \\[4pt]
\alpha_0 + (1-\alpha_0)\dfrac{it-it_{\mathrm{stage1,end}}}{N_{\mathrm{warm}}},
& it_{\mathrm{stage1,end}} \le it < it_{\mathrm{stage2,end}}, \\[8pt]
1, & it \ge it_{\mathrm{stage2,end}},
\end{cases}
\label{eq:alpha_warmup}
\end{equation}
where $it$ denotes the current optimization iteration, $it_{\mathrm{step1,end}}$ is the iteration at which Stage~1 ends, and $it_{\mathrm{stage2,end}}$ is the iteration at which Stage~2 ends. Here, $\alpha_0$ is the initial warmup value at the beginning of Stage~2, and $N_{\mathrm{warm}}$ denotes the effective number of iteration intervals in the warmup stage.
    
For the $180^\circ$ training case, $l$ is fixed to the value identified in Sta~1, while the coupled parameters $G$ and $a_4$ are gradually activated. For $90^\circ$ training, $\mu$ is fixed and $a_3$ is gradually activated.

\paragraph{Stage 3 (full release and GradNorm-based joint refinement)}
In the final stage, all intended trainable quantities are released and the full loss function is activated. The remaining physics-based constraints are introduced/strengthened, and adaptive multi-objective balancing (GradNorm) is applied to prevent the optimization from being dominated by a single loss component. This stage performs the final joint refinement of the material parameters and the predicted field distributions under the complete physics-informed formulation.

\section{Identification of the $90^\circ$ Domain Wall Switching Path from MD Data}
\label{app:90dw_switching_path}

\setcounter{figure}{0}
\setcounter{table}{0}
\renewcommand{\thefigure}{\thesection\arabic{figure}}
\renewcommand{\thetable}{\thesection\arabic{table}}

To identify a representative transition-state polarization for the stationary-point constraint, we post-processed the relaxed $90^\circ$ DW configuration obtained from MD. Since the MD output is given in the global coordinate system and local polarization vectors are available only at Ti sites, we retained the Ti-centered polarization data and transformed the polarization vectors to the crystal coordinate system. The transformed data were then sorted along the wall-normal direction and averaged within one-dimensional bins to construct the profiles $P_x(x)$, $P_y(x)$, and $P_z(x)$.

\begin{figure}[H]
    \centering

    \begin{subfigure}[b]{0.45\textwidth}
        \centering
        \includegraphics[width=\linewidth]
        {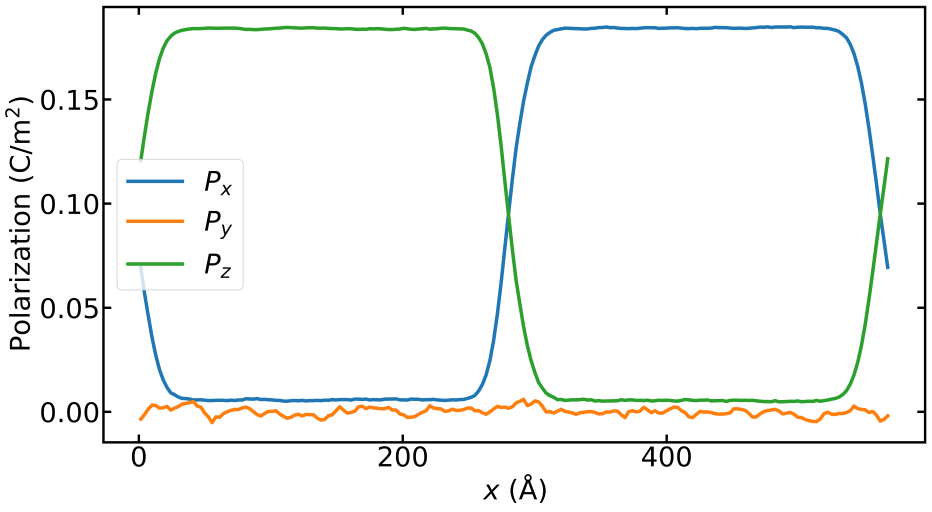}
        \caption{}
    \end{subfigure}
    \hspace{0.01\textwidth}
    \begin{subfigure}[b]{0.3\textwidth}
        \centering
        \includegraphics[width=\linewidth]
        {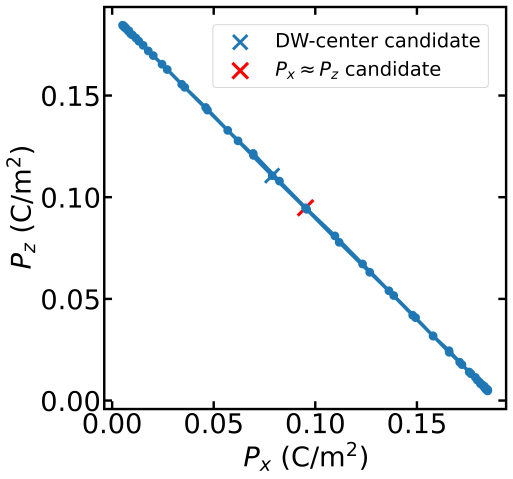}
        \caption{}
    \end{subfigure}
  
    \caption{
    (a) one-dimensional profiles of $90^\circ$ DW in crystal coordinate system, 
    (b) reconstructed switching path in polarization space.}
    \label{fig:app_90dw_switching_path}
\end{figure}

Figure~\ref{fig:app_90dw_switching_path}(a) shows that $P_x$ and $P_z$ vary complementarily across the wall, whereas $P_y$ remains close to zero, indicating that the polarization rotation occurs predominantly in the $(P_x,P_z)$ plane. The corresponding switching path reconstructed in polarization space is shown in Fig.~\ref{fig:app_90dw_switching_path}(b), where each point represents one bin-averaged polarization state projected onto the $(P_x,P_z)$ plane. The representative transition-state candidate is identified as the point where \(P_x\) and \(P_z\) are closest in magnitude. Its polarization is \((P_x,\,P_y,\,P_z)=(0.0952,\,-0.0044,\,0.0948)\ \mathrm{C/m^2}\). After normalization by the spontaneous polarization magnitude \(P_s=0.18\ \mathrm{C/m^2}\), the transition-state polarization becomes \(\mathbf{P}^{\mathrm{ts}}\approx(0.529,\,-0.024,\,0.527)\). To ensure that the transition state respects the corresponding crystal symmetries, the small out-of-plane \(P_y\) component is set to zero and we further symmetrize the in-plane components by setting \(P_x=P_z\), yielding the simplified transition-state polarization adopted in the training, \(\mathbf{P}^{\mathrm{ts}}\approx(0.528,\,0,\,0.528)\).

\section*{Data availability}
We provide free access to the Python code under the URL:
\url{https://github.com/xwang1994/phasefield-pinn-ferroelectrics-180dw}

\bibliographystyle{elsarticle-num}
\bibliography{references}

@article{durdiev2025parameterization,
  title={Parameterization of a phase field model for ferroelectrics from molecular dynamics data},
  author={Durdiev, Dilshod and Wendler, Frank and Zaiser, Michael and Azuma, Hikaru and Tsuzuki, Takahiro and Ogata, Shuji and Ogawa, Tomohiro and Kobayashi, Ryo and Uranagase, Masayuki},
  journal={Acta Mater.},
  volume={283},
  pages={120513},
  year={2025},
  doi={10.1016/j.actamat.2024.120513}
}

@article{azuma2025unique,
  title={Unique temperature-dependence of polarization switching paths in ferroelectric BaTiO\(_3\): A molecular dynamics simulation study},
  author={Azuma, Hikaru and Ogawa, Tomohiro and Ogata, Shuji and Kobayashi, Ryo and Uranagase, Masayuki and Tsuzuki, Takahiro and Wendler, Frank},
  journal={Acta Mater.},
  volume={296},
  pages={121216},
  year={2025},
  doi={10.1016/j.actamat.2025.121216}
}

@article{raissi2020hidden,
  title={Hidden fluid mechanics: Learning velocity and pressure fields from flow visualizations},
  author={Raissi, Maziar and Yazdani, Alireza and Karniadakis, George Em},
  journal={Science},
  volume={367},
  number={6481},
  pages={1026--1030},
  year={2020},
  doi={10.1126/science.aaw4741}
}

@article{azuma2023microscopic,
  title={Microscopic structure and migration of 90\degree{} ferroelectric domain wall in {BaTiO\(_3\)} determined via molecular dynamics simulations},
  author={Azuma, Hikaru and Ogata, Shuji and Kobayashi, Ryo and Uranagase, Masayuki and Tsuzuki, Takahiro and Durdiev, Dilshod and Wendler, Frank},
  journal={J. Appl. Phys.},
  volume={133},
  number={10},
  pages={104101},
  year={2023},
  doi={10.1063/5.0138332}
}

@article{schrade2014invariant,
  title={An invariant formulation for phase field models in ferroelectrics},
  author={Schrade, David and Müller, Ralf and Gross, Dietmar and Keip, M-A and Thai, H and Schröder, J},
  journal={Int. J. Solids Struct.},
  volume={51},
  number={11--12},
  pages={2144--2156},
  year={2014}
}

@article{schrade2007domain,
  title={Domain evolution in ferroelectric materials: A continuum phase field model and finite element implementation},
  author={Schrade, D. and Mueller, R. and Xu, B.X. and Gross, D.},
  journal={Computer Methods in Applied Mechanics and Engineering},
  volume={196},
  number={41--44},
  pages={4365--4374},
  year={2007},
  publisher={Elsevier},
  doi={10.1016/j.cma.2007.05.010}
}

@article{matsuo2023,
  author={Matsuo, Hiroki},
  title={Domain-wall photovoltaic effect in ferroelectric perovskite oxides},
  journal={J. Ceram. Soc. Jpn.},
  volume={131},
  number={8},
  pages={429--436},
  year={2023},
  doi={10.2109/jcersj2.23084}
}

@article{tao2024,
  title={Simultaneously enhanced electrical properties and high-power characteristics of {(K,Na)NbO$_3$} lead-free piezoceramics by hot-pressing},
  author={Tao, Chuanyang and Chen, Binjie and Lin, Tao and Ma, Jun and Dou, Zhongshang and Zhang, Mao-Hua and Zhong, Meipeng and Gong, Wen and Zhou, Yuqing and Yao, Fang-Zhou and Wang, Ke},
  journal={Ceram. Int.},
  volume={50},
  pages={28047--28053},
  year={2024}
}

@article{zhao2022coherent,
  title={Coherent Precipitates with Strong Domain Wall Pinning in Alkaline Niobate Ferroelectrics},
  author={Zhao, Changhao and Gao, Shuang and Kleebe, Hans-Joachim and Tan, Xiaoli and Koruza, Jurij and Rödel, Jürgen},
  journal={Adv. Mater.},
  volume={34},
  number={34},
  pages={2202379},
  year={2022},
  doi={10.1002/adma.202202379}
}

@article{zhao2021precipitation,
  title={Precipitation Hardening in Ferroelectric Ceramics},
  author={Zhao, Changhao and Gao, Shuang and Yang, Tiannan and Scherer, Michael and Schultheiß, Jan and Meier, Dennis and Tan, Xiaoli and Kleebe, Hans-Joachim and Chen, Long-Qing and Koruza, Jurij and Rödel, Jürgen},
  journal={Adv. Mater.},
  year={2021},
  doi={10.1002/adma.202102421}
}

@article{gao2023topology,
  title={Precipitate-domain wall topologies in hardened Li-doped NaNbO\(_3\)},
  author={Gao, Shuang and Zhao, Changhao and Bohnen, Matthias and Müller, Ralf and Rödel, Jürgen and Kleebe, Hans-Joachim},
  journal={Acta Mater.},
  volume={254},
  pages={119999},
  year={2023},
  doi={10.1016/j.actamat.2023.119999}
}

@article{hofling2021control,
  title={Control of polarization in bulk ferroelectrics by mechanical dislocation imprint},
  author={Höfling, Marion and Zhou, Xiandong and Reimer, Lukas M and Bruder, Enrico and Liu, Binzhi and Zhou, Lin and Groszewicz, Pedro B and Zhuo, Fangping and Rödel, Jürgen},
  journal={Science},
  volume={372},
  number={6545},
  pages={961--964},
  year={2021},
  doi={10.1126/science.abe3810}
}

@article{maguire2024direct,
  title={Direct Imaging of Built-In Electric Fields at Ferroelectric Domain Walls by High-Voltage Kelvin Probe Force Microscopy},
  author={Maguire, J and Dwyer, C and Jesse, S and Kalinin, S V},
  journal={Nano Lett.},
  volume={24},
  number={1},
  pages={123--130},
  year={2024},
  doi={10.1021/acs.nanolett.3c02966}
}

@article{doherty2023domain,
  title={Domain Wall Nanoelectronics in Ferroelectrics},
  author={Doherty, J P and Balke, N and Kalinin, S V and Gregg, J M},
  journal={Nat. Phys.},
  volume={19},
  number={4},
  pages={441--451},
  year={2023},
  doi={10.1038/s41567-022-01921-4}
}

@article{hadjimichael2018domain,
  title={Domain Wall Engineering and Topology in Ferroelectric Nanostructures Revealed by Synchrotron X-ray Nanodiffraction},
  author={Hadjimichael, M and Nord, M and Vrejoiu, I and Zschornak, M and Pietsch, U and Schmidbauer, M},
  journal={Phys. Rev. Lett.},
  volume={120},
  number={3},
  pages={037602},
  year={2018},
  doi={10.1103/PhysRevLett.120.037602}
}

@article{raissi2019physics,
  title={Physics-informed neural networks: A deep learning framework for solving forward and inverse problems involving nonlinear partial differential equations},
  author={Raissi, Maziar and Perdikaris, Paris and Karniadakis, George Em},
  journal={J. Comput. Phys.},
  volume={378},
  pages={686--707},
  year={2019}
}

@article{karniadakis2021physics,
  title={Physics-informed machine learning},
  author={Karniadakis, George Em and Kevrekidis, Ioannis G and Lu, Lu and Perdikaris, Paris and Wang, Sifan and Yang, Liu},
  journal={Nat. Rev. Phys.},
  volume={3},
  number={6},
  pages={422--440},
  year={2021}
}

@article{cuomo2022scientific,
  title={Scientific machine learning through physics-informed neural networks: Where we are and what's next},
  author={Cuomo, Salvatore and Di Cola, Vittorio Salvatore and Giampaolo, Francesco and Rozza, Gianluigi and Raissi, Maziar and Piccialli, Francesco},
  journal={J. Sci. Comput.},
  volume={92},
  number={3},
  pages={1--38},
  year={2022}
}

@article{shang2024quantification,
  title={Quantification of gradient energy coefficients using physics-informed neural networks},
  author={Shang, Lan and Zhao, Yunhong and Zheng, Sizheng and Wang, Jin and Zhang, Tongyi and Wang, Jie},
  journal={Int. J. Mech. Sci.},
  volume={273},
  pages={109210},
  year={2024},
  doi={10.1016/j.ijmecsci.2023.109210}
}

@article{zhang2024te,
  title={Room-temperature ferroelectric, piezoelectric and resistive switching behaviors of single-element Te nanowires},
  author={Zhang, Jinlei and Zhang, Jiayong and Qi, Yaping and Gong, Shuainan and Xu, Hang and Liu, Zhenqi and Zhang, Ran and Sadi, Mohammad A and Sychev, Demid and Zhao, Run and Yang, Hongbin and Wu, Zhenping and Cui, Dapeng and Wang, Lin and Ma, Chunlan and Wu, Xiaoshan and Gao, Ju and Chen, Yong P and Wang, Xinran and Jiang, Yucheng},
  journal={Nat. Commun.},
  volume={15},
  pages={7648},
  year={2024},
  doi={10.1038/s41467-024-52062-6}
}

@article{zhao2025why,
  title={Why not inorganic ferroelectrics: Harnessing the pyroelectric charges for triboelectric nanogenerators},
  author={Zhao, Pengfei and Huang, Yue and Li, Pengfei and Sharma, Niyorjyoti and Murdoch, Billy J and Rogers, Andrew and Wang, Chia-Hsin and Sharma, Surbhi and Wu, Zhizheng and Chen, Jinkai and Sun, Tao and Lei, Jintao and Kumar, Amit and Soin, Navneet},
  journal={Nano Energy},
  year={2025},
  doi={10.1016/j.nanoen.2025.111046}
}

@article{thapa2024microsecond,
  title={Microsecond electro-optic switching in the nematic phase of a ferroelectric nematic liquid crystal},
  author={Thapa, Kamal and Paladugu, Sathyanarayana and Lavrentovich, Oleg D},
  journal={Opt. Express},
  volume={32},
  number={23},
  pages={40274--40286},
  year={2024},
  doi={10.1364/OE.541317}
}

@article{zhao2024dielectric,
  title={Dielectric nonlinearity analysis of {BNT--ST--BT} relaxor ferroelectric thin films with different film thicknesses},
  author={Zhao, Jinyan and Wang, Zhe and Li, Yizhuo and Zheng, Kun and Zhang, Jie and Meng, Haoyan and Zhang, Nan and Zhao, Yulong and Niu, Gang and Ren, Wei},
  journal={J. Appl. Phys.},
  volume={136},
  number={22},
  pages={224103},
  year={2024},
  doi={10.1063/5.0231329}
}

@article{liu2025flexible,
  title={A {BaTiO$_3$}-based flexible ferroelectric capacitor for non-volatile memories},
  author={Liu, Xingpeng and Wei, Chunshu and Sun, Tangyou and Zhang, Fabi and Li, Haiou and Liu, Linsheng and Peng, Ying and Li, Hezhang and Hong, Min},
  journal={J. Materiomics},
  volume={11},
  number={2},
  pages={100870},
  year={2025},
  doi={10.1016/j.jmat.2024.04.001}
}

@article{liu2025superior,
  title={Superior energy-storage performance in {BaTiO\(_3\)}--{AgNbO\(_3\)} binary relaxor via the competitions of multiple polar orders},
  author={Liu, Minghao and Liu, Hongbo and Liu, Zhen and Hu, Zimeng and Dai, Kai and Yan, Shiguang and Hu, Zhigao and Wang, Genshui},
  journal={Acta Mater.},
  volume={289},
  pages={120943},
  year={2025},
  doi={10.1016/j.actamat.2025.120943}
}

@article{yang2025effects,
  title={Effects of trace {Nb} dopant on core-shell microstructure and ferroelectric domain switching in {BiFeO\(_3\)}--{BaTiO\(_3\)} ceramics},
  author={Yang, Ziqi and Li, Yizhe and Pan, Juncheng and Xie, Bingying and Li, Kexue and Moore, Katie L and Kleppe, Annette K and Hall, David A},
  journal={Acta Mater.},
  volume={289},
  pages={120890},
  year={2025},
  doi={10.1016/j.actamat.2025.120890}
}

@article{liu2021quadrupole,
  title={Engineered periodic quadrupole superstructure in ferroelectric thin films via flexoelectricity},
  author={Liu, Zhen and Zhou, Xiandong and Feng, Biao and Xu, Bai-Xiang},
  journal={Acta Mater.},
  volume={216},
  pages={117126},
  year={2021},
  doi={10.1016/j.actamat.2021.117126}
}

@article{zhou2025cooling,
  title={The influence of cooling rates on strain phase diagrams and domain structures of ferroelectric thin films: A case study of {PbTiO\(_3\)}},
  author={Zhou, Meng-Jun and Zhang, Peng and Wang, Bo and Yi, Di and Nan, Ce-Wen},
  journal={Acta Mater.},
  volume={296},
  pages={121207},
  year={2025},
  doi={10.1016/j.actamat.2025.121207}
}

@article{mi2021breakdown,
  title={Phase field modeling of dielectric breakdown of ferroelectric polymers subjected to mechanical and electrical loadings},
  author={Mi, Zhang and Zhang, Yong and Hou, Xu and Wang, Jie},
  journal={Int. J. Solids Struct.},
  volume={217},
  number={218},
  pages={123--133},
  year={2021},
  doi={10.1016/j.ijsolstr.2021.02.009}
}

@article{liu2022vortex,
  title={Phase-field simulations of vortex chirality manipulation in ferroelectric thin films},
  author={Liu, Di and Wang, Jing and Jafri, Hasnain Mehdi and Wang, Xueyun and Shi, Xiaoming and Liang, Deshan and Yang, Chao and Cheng, Xingwang and Huang, Houbing},
  journal={NPJ Quantum Mater.},
  volume={7},
  number={34},
  year={2022},
  doi={10.1038/s41535-022-00444-8}
}

@article{Zhu2024HfO2Review,
  author={Zhu, Tianyuan and Ma, Liyang and Deng, Shiqing and Liu, Shi},
  title={Progress in computational understanding of ferroelectric mechanisms in {HfO\(_2\)}},
  journal={npj Comput. Mater.},
  volume={10},
  number={1},
  pages={88},
  year={2024},
  doi={10.1038/s41524-024-01352-0}
}

@article{Ali2025PbTiO3Adsorption,
  author={Ali, Ijaz and Liu, Jian-An and Yin, Li-Chang and Wang, Lianzhou and Liu, Gang},
  title={Water adsorption on ferroelectric PbTiO\(_3\) (001) surface: A density functional theory study},
  journal={J. Colloid Interface Sci.},
  volume={678},
  pages={984--991},
  year={2025},
  doi={10.1016/j.jcis.2024.09.079}
}

@article{gazis1963elastic,
  title={The elastic tensor of given symmetry nearest to an anisotropic elastic tensor},
  author={Gazis, D C and Tadjbakhsh, I and Toupin, R A},
  journal={Acta Crystallogr.},
  volume={16},
  pages={917--922},
  year={1963},
  doi={10.1107/S0365110X63002431}
}

@article{moakher2006closest,
  title={The closest elastic tensor of arbitrary symmetry to an elasticity tensor of lower symmetry},
  author={Moakher, Maher and Norris, Andrew N},
  journal={J. Elasticity},
  volume={85},
  number={3},
  pages={215--263},
  year={2006},
  doi={10.1007/s10659-006-9082-0}
}

@article{ranganathan2008universal,
  title={Universal elastic anisotropy index},
  author={Ranganathan, Shivakumar I and Ostoja-Starzewski, Martin},
  journal={Phys. Rev. Lett.},
  volume={101},
  number={5},
  pages={055504},
  year={2008},
  doi={10.1103/PhysRevLett.101.055504}
}

@article{Catalan2012DomainWallNanoelectronics,
  title={Domain wall nanoelectronics},
  author={Catalan, Gustau and Seidel, Jan and Ramesh, Ramamoorthy and Scott, James F},
  journal={Rev. Mod. Phys.},
  volume={84},
  number={1},
  pages={119--156},
  year={2012},
  doi={10.1103/RevModPhys.84.119}
}

@article{Nataf2020DomainWallEngineering,
  author={Nataf, Guillaume F and Guennou, Mael and Gregg, J Marty and Meier, Denis and Hlinka, Jiri and Salje, Ekhard K H and Kreisel, J{\"o}rg},
  title={Domain-wall engineering and topological defects in ferroelectric and ferroelastic materials},
  journal={Nat. Rev. Phys.},
  volume={2},
  number={11},
  pages={634--648},
  year={2020},
  doi={10.1038/s42254-020-0235-z}
}

@article{rojas2023parameter,
  title={Parameter identification for a damage phase field model using a physics-informed neural network},
  author={Rojas, Carlos J G and Boldrini, Jos L and Bittencourt, Marco L},
  journal={Theor. Appl. Mech. Lett.},
  volume={13},
  pages={100450},
  year={2023},
  doi={10.1016/j.taml.2023.100450}
}

@article{Hashimoto2015MDpermittivity,
  author  = {Hashimoto, T. and Moriwake, H.},
  title   = {Dielectric properties of {BaTiO$_3$} by molecular dynamics simulations using a shell model},
  journal = {Molecular Simulation},
  volume  = {41},
  number  = {13},
  pages   = {1074--1080},
  year    = {2015}
}

@article{Wang2010BTO_JAP,
  author    = {Wang, J. J. and Zhang, T. Y.},
  title = {Lattice, elastic, polarization, and electrostrictive properties of {BaTiO$_3$} from first principles},
  journal   = {Journal of Applied Physics},
  volume    = {108},
  number    = {3},
  pages     = {034107},
  year      = {2010},
  publisher = {AIP Publishing},
  doi       = {10.1063/1.3457363}
}

@article{Pandech2015ATiO3,
  author    = {Pandech, N. and Bovornratanaraks, T. and Usui, H. and Oshiyama, A.},
  title = {Elastic properties of perovskite {ATiO$_3$} ({A} = {Be}, {Mg}, {Ca}, {Sr}, and {Ba}) and {PbBO$_3$} from first principles},
  journal   = {Journal of Applied Physics},
  volume    = {117},
  number    = {17},
  pages     = {174108},
  year      = {2015},
  publisher = {AIP Publishing},
  doi       = {10.1063/1.4919775}
}

@article{Sakhya2015ATiO3,
  author    = {Sakhya, A. P. and Choudhary, R. N. P. and Das, B.},
  title     = {Electronic structure and elastic properties of {ATiO$_3$} (A = {Ba}, {Sr}, {Ca}) perovskites: A first-principles study},
  journal   = {Indian Journal of Pure and Applied Physics},
  volume    = {53},
  number    = {5},
  pages     = {331--339},
  year      = {2015}
}

@article{Choithrani2014BTOElastic,
  author    = {Choithrani, R. and Sharma, R. and Singh, M.},
  title     = {Structural, elastic and thermal properties of {BaTiO$_3$}},
  journal   = {Central European Journal of Physics},
  volume    = {12},
  number    = {11},
  pages     = {759--767},
  year      = {2014},
  doi       = {10.2478/s11534-014-0532-5},
  publisher = {Springer}
}

@article{kingma2015adam,
  title   = {Adam: A Method for Stochastic Optimization},
  author  = {Kingma, Diederik P. and Ba, Jimmy},
  journal = {International Conference on Learning Representations},
  year    = {2015},
  doi     = {10.48550/arXiv.1412.6980}
}

@article{chen2018gradnorm,
  title={GradNorm: Gradient Normalization for Adaptive Loss Balancing in Deep Multitask Networks},
  author={Chen, Zhao and Badrinarayanan, Vijay and Lee, Chen-Yu and Rabinovich, Andrew},
  journal={Proc. Int. Conf. Mach. Learn. (ICML)},
  volume={80},
  pages={794--803},
  year={2018}
}

@article{Kumar2023FerroX,
  author    = {Kumar, Prabhat and Nonaka, Andrew and Jambunathan, Revathi and Pahwa, Girish and Salahuddin, Sayeef and Jackie, Zhi Yao},
  title     = {FerroX: A GPU-accelerated, 3D phase-field simulation framework for modeling ferroelectric devices},
  journal   = {Computer Physics Communications},
  volume    = {290},
  pages     = {108757},
  year      = {2023},
  doi       = {10.1016/j.cpc.2023.108757},
  publisher = {Elsevier}
}

@article{Liu2022VortexChirality,
  author    = {Liu, Di and Wang, Jing and Jafri, Hasnain Mehdi and Wang, Xueyun and Shi, Xiaoming and Liang, Deshan and Yang, Chao and Cheng, Xingwang and Huang, Houbing},
  title     = {Phase-field simulations of vortex chirality manipulation in ferroelectric thin films},
  journal   = {npj Quantum Materials},
  volume    = {7},
  pages     = {34},
  year      = {2022},
  doi       = {10.1038/s41535-022-00434-6},
  publisher = {Nature Publishing Group}
}

@article{Zhou2022DislocationDW,
  author    = {Zhou, Xiandong and Liu, Zhen and Xu, Bai-Xiang},
  title     = {Influence of dislocations on domain walls in perovskite ferroelectrics: Phase-field simulation and driving force calculation},
  journal   = {International Journal of Solids and Structures},
  volume    = {238},
  pages     = {111391},
  year      = {2022},
  doi       = {10.1016/j.ijsolstr.2021.111391},
  publisher = {Elsevier}
}

@article{Durdiev2023FourierPF,
  author    = {Durdiev, Dilshod and Wendler, Frank},
  title     = {An effective Fourier spectral phase-field approach for ferroelectric materials},
  journal   = {Computational Materials Science},
  volume    = {218},
  pages     = {111928},
  year      = {2023},
  doi       = {10.1016/j.commatsci.2022.111928},
  publisher = {Elsevier}
}

@article{Cochran1960Ferroelectricity,
  author  = {W. Cochran},
  title   = {Crystal Stability and the Theory of Ferroelectricity},
  journal = {Physical Review Letters},
  volume  = {3},
  pages   = {412--414},
  year    = {1960}
}

@article{Dieguez2022TranslationalCovarianceFlexoelectricity,
  author  = {Oswaldo Di{\'e}guez and Massimiliano Stengel},
  title   = {Translational covariance of flexoelectricity at ferroelectric domain walls},
  journal = {arXiv preprint},
  volume  = {arXiv:2201.12561v2},
  pages   = {1--19},
  year    = {2022}
}

@article{Shu2001DomainPatternsMacroscopic,
  author  = {Y. C. Shu and K. Bhattacharya},
  title   = {Domain patterns and macroscopic behaviour of ferroelectric materials},
  journal = {Philosophical Magazine B},
  volume  = {81},
  number  = {12},
  pages   = {2021--2054},
  year    = {2001},
  doi     = {10.1080/13642810108208556}
}

@inproceedings{Guo2017Calibration,
  author    = {Chuan Guo and Geoff Pleiss and Yu Sun and Kilian Q. Weinberger},
  title     = {On Calibration of Modern Neural Networks},
  booktitle = {Proceedings of the 34th International Conference on Machine Learning},
  series    = {Proceedings of Machine Learning Research},
  volume    = {70},
  pages     = {1321--1330},
  year      = {2017},
  publisher = {PMLR}
}

@article{Huang2022HomPINNs,
  author    = {Yao Huang and Wenrui Hao and Guang Lin},
  title     = {HomPINNs: Homotopy physics-informed neural networks for learning multiple solutions of nonlinear elliptic differential equations},
  journal   = {Computers and Mathematics with Applications},
  volume    = {121},
  pages     = {62--73},
  year      = {2022},
  publisher = {Elsevier}
}

@inproceedings{Krishnapriyan2021FailureModesPINN,
  author    = {Aditi S. Krishnapriyan and Amir Gholami and Shandian Zhe and Robert M. Kirby and Michael W. Mahoney},
  title     = {Characterizing possible failure modes in physics-informed neural networks},
  booktitle = {Advances in Neural Information Processing Systems},
  year      = {2021},
  publisher = {NeurIPS}
}

@article{Dwivedi2025PIELM,
  author    = {Vikas Dwivedi and Bruno Sixou and Monica Sigovan},
  title     = {Curriculum learning-driven PIELMs for fluid flow simulations},
  journal   = {Neurocomputing},
  volume    = {650},
  pages     = {130924},
  year      = {2025},
  publisher = {Elsevier}
}

@article{Chen2025PINNJoint,
  title   = {PINN-based joint identification and low-dimensional dynamical modeling of joint-assembled structures},
  author  = {Chen, Chao and Wang, Yilong and Zhang, Xiaoyun and Chen, Shuai and Fang, Bo and Chen, Wanying and Cao, Dengqing and Han, Hesheng},
  journal = {International Journal of Mechanical Sciences},
  year    = {2025},
  doi     = {10.1016/j.ijmecsci.2025.111109}
}

@article{Shang2024GradientPINN,
  title   = {Quantification of gradient energy coefficients using physics-informed neural networks},
  author  = {Shang, Lan and Zhao, Yunhong and Zheng, Siheng and Wang, Jin and Zhang, Tongyi and Wang, Jie},
  journal = {International Journal of Mechanical Sciences},
  volume  = {273},
  year    = {2024},
  month   = {July},
  pages   = {109210},
  doi     = {10.1016/j.ijmecsci.2024.109210}
}

@article{Zhou2026ElectromechanicalNBT,
  title   = {Analytical and phase-field study of anomalous electromechanical behaviors in {Na$_{0.5}$Bi$_{0.5}$TiO$_3$-xSrTiO$_3$}},
  author  = {Zhou, Yucheng and Wang, Kai and Chen, Wanxin and Zhang, Chunli and Huang, Kefu and Wang, Shuai},
  journal = {International Journal of Mechanical Sciences},
  year    = {2026},
  doi     = {10.1016/j.ijmecsci.2026.111531}
}

@article{Chen2023TopologicalPhononic,
  title   = {Temperature-controlled elastic wave transport in topological ferroelectric phononic crystal plates},
  author  = {Chen, Zhenyu and Zhou, Weijian},
  journal = {International Journal of Mechanical Sciences},
  volume  = {241},
  year    = {2023},
  month   = {March},
  pages   = {107964},
  doi     = {10.1016/j.ijmecsci.2022.107964}
}

@article{zheng2025chiral,
  title={Origin of Chiral Phase Transition of Polar Vortex in Ferroelectric/Dielectric Superlattices},
  author={Zheng, Sizheng and Zhang, Jingtong and Li, Ailin and Wang, Jie},
  journal={Nano Letters},
  volume={25},
  number={4},
  pages={xxxx--xxxx},
  year={2025},
  doi={10.1021/acs.nanolett.xxxxx}
}
\end{document}